\documentclass[seceq,preprint]{ptptex}
\usepackage{amsmath,amssymb,comment}
\usepackage{graphicx}
\usepackage{graphicx, color}
 \notypesetlogo
 \def\ind{\indent}
 \def\nn{\nonumber}
 
 \def\beq{\begin{equation}}
 \def\eeq{\end{equation}}
 \def\ben{\begin{enumerate}}
 \def\een{\end{enumerate}}
 \def\ble{\begin{flushleft}}
 \def\ele{\end{flushleft}}
 \def\btab{\begin{tabular}}
 \def\etab{\end{tabular}}
 
 \def\bri{\begin{flushright}}
 \def\eri{\end{flushright}}
 \def\bce{\begin{center}}
 \def\ece{\end{center}}
 \def\beqa{\begin{eqnarray}}
 \def\eeqa{\end{eqnarray}}
 \def\bea*{\begin{eqnarray*}}
 \def\eea*{\end{eqnarray*}}
 \def\barr{\begin{array}}
 \def\earr{\end{array}}
 \def\bit{\begin{itemize}}
 \def\eit{\end{itemize}}
 
 \def\cL{\mbox{${\cal L}$}}

\def\cL{\mbox{${\cal L}$}}

\def\cP{\mbox{${\cal P}$}}

\def\be{\mbox{\boldmath $e$}}

\def\bi{\mbox{\boldmath $i$}}
\def\bj{\mbox{\boldmath $j$}}
\def\bk{\mbox{\boldmath $k$}}

\def\bn{\mbox{\boldmath $n$}}

\def\bp{\mbox{\boldmath $p$}}

\def\bx{\mbox{\boldmath $x$}}

\def\nn{\nonumber}
\def\dslash{\partial{\raise 1pt\hbox{$\!\!\!/$}}}

 \def\ind{\indent}
\def\underr#1{$\underline{\smash{\raise 0.5ex\hbox{#1}}}$}
  \def\under#1{$\underline{\smash{\raise 0.1ex\hbox{#1}}}$}

\title{%
Quaternions, Lorentz Group and
the Dirac Theory
}
\vspace{5mm}
\author{%
  Katsusada {\sc Morita}
}
\inst{%
{\it  Department of Physics, 
  Nagoya University, Nagoya 464-4602, Japan}
}
\recdate{%
January 5, 2007
}
\vspace{5mm}
\abst{%
It is shown that
a subgroup of $SL(2,{\mathbb H})$,
denoted $Spin(2,{\mathbb H})$ in this paper,
which is defined by two conditions
in addition to unit quaternionic determinant,
is locally isomorphic to
the restricted Lorentz group, $L_+^\uparrow$.
On the basis of the Dirac theory 
using the spinor group $Spin(2,{\mathbb H})$,
in which
the charge conjugation transformation
becomes linear in the quaternionic Dirac spinor,
it is shown that the
Hermiticity requirement
of the Dirac Lagrangian,
together with the persistent
presence of the Pauli-G\"ursey $SU(2)$
group, requires
an additional imaginary unit (taken
to be the ordinary one, $i$)
that commutes
with Hamilton's units,
in the theory.
A second quantization
is performed with this $i$
incorporated into the theory,
and we recover
the conventional Dirac theory
with an automatic `anti-symmetrization'
of the field operators.
It is also pointed out that
we are  naturally
led to the scheme of complex quaternions,
${\mathbb H}^c$,
in which a space-time point is represented by
a Hermitian quaternion,
and that
the isomorphism
$SL(1,{\mathbb H}^c)/Z_2\cong L_+^\uparrow$ 
is a direct consequence of the
fact $Spin(2,{\mathbb H})/Z_2\cong L_+^\uparrow$.
Using $SL(1,{\mathbb H}^c)\cong SL(2,{\mathbb C})$,
we make explicit the Weyl spinor indices
of the spinor-quaternion,
which is the Dirac spinor defined over ${\mathbb H}^c$.
}
\begin{document}
\baselineskip=15.7pt
\maketitle
\section{Introduction}                                          %
In the past years,
there have been many attempts
to apply quaternions \cite{1)}
to relativity, despite the essential
difference that
the division algebra, ${\mathbb H}$, of
real quaternions,
which occupies a unique position
in mathematics, can be regarded as the
4-dimensional Euclidean space ${\mathbb R}^4$,
while the Minkowski space-time is a
4-dimensional pseudo-Euclidean space
with signature 2.
This difference
has been overcome
by employing complex quaternions,
suggested by the introduction
of an imaginary time 
coordinate. \cite{2)}
However, in relativity, $i$ is completely
superfluous.\footnote{Misner, Thorne and Wheeler
declared a farewell to ``$x^4=ict$'' from
the viewpoint of general relativity. (See Box 2.1 in their book
{\it Gravitation}, 1970.)} 
It is more desirable
to represent the Lorentz transformations
in the quaternionic approach
without recourse to $i$.
Since it is obvious that
the space-time point cannot be written
as a real quaternion,
we should consider the next simplest case, a
$2\times 2$ quaternionic matrix,
which in turn implies a 2-component
quaternionic spinor.
The purpose of this
paper
is to show
that
the Lorentz transformations
can be represented in terms of
Hamilton's real quaternions only
in a manner similar to that used in
spinor analysis,
while
a consistent Lagrangian formulation
of the Dirac theory based on
Hamilton's real quaternions
requires an $i$ that
commutes with them.
\\[2mm]
\ind
The spinor analysis
is based on the field of complex numbers.
Spinor analysis
over the field
of real quaternions
involves $SL(2,{\mathbb H})$ spinors,
which are Weyl spinors in $D=6$
space-time,
and it was studied thoroughly by Kugo and Townsend \cite{3)}
in relation to $D=6$ supersymmetry. However,
this has no immediate application
to 4-dimensional Minkowski space-time,
because $SL(2,{\mathbb H})$ is locally isomorphic
to the conformal group $SO(5,1)$ in ${\mathbb R}^4$.
As is well known,
the Lorentz group
$SO(3,1)$ is a subgroup of $SO(5,1)$.
In this paper, we seek a spinor analysis
over the field
of real quaternions,
so that
we end up with groups that are
no wider or narrower than
the Lorentz group $SO(3,1)$.
\\[2mm]
\ind
To this end recall that,
in the spinor analysis,
the restricted Lorentz transformations
are represented by 2-dimensional
complex unimodular matrices
which form the group $SL(2,{\mathbb C})$.
Thus,
we need Pauli matrices including
the unit matrix $\sigma_0$.
The set of
three Pauli matrices
$\{\sigma_1,\sigma_2,\sigma_3\}$
furnishes an irreducible representation
of an odd Clifford algebra of rank 1.
It is therefore natural
to associate three spatial coordinates
with them.
That is,
an arbitrary space-time point
$(x^0,x^1,x^2,x^3)$
is represented
by a complex Hermitian matrix of the form
\beqa
X&=&x^0\sigma_0+x^1\sigma_1+x^2\sigma_2+x^3\sigma_3\nn\\[2mm]
&=&x^0
\left(
\barr{cc}
1&0\\
0&1\\
\earr
\right)
+
x^1
\left(
\barr{cc}
0&1\\
1&0\\
\earr
\right)
+
x^2
\left(
\barr{cc}
0&-i\\
i&0\\
\earr
\right)
+
x^3
\left(
\barr{cc}
1&0\\
0&-1\\
\earr
\right),
\label{eqn:1-1}
\eeqa
whose determinant $(x^0)^2-(x^1)^2-(x^2)^2-(x^3)^2$
is Lorentz invariant.
The restricted Lorentz transfomation
is then represented by $A\in SL(2,{\mathbb C})$
through
$X\to X'=AXA^\dag$,
which leaves the determinant invariant, i.e.
det$\,\,X'=$det$\,\,X$.
\\[2mm]
\ind
It should be noted, however, that
only the second Pauli matrix, $\sigma_2$,
contains the imaginary unit $i$.
Since
there exists only one imaginary
unit in complex numbers,
any complex Hermitian matrix
can be written in the form
(\ref{eqn:1-1})
for real $x^i\;(i=0,1,2,3)$.
By contrast,
quaternions
have three imaginary units (Hamilton's units),
$\bi,\bj$ and $\bk$,
the same number as the dimensionality of our space.
It is tempting,
therefore,
to assume that three spatial coordinates
are formally associated
with the three matrices
obtained
by replacing $i$ in $\sigma_2$ with
$\bi,\bj,\bk$.
This implies that we associate an arbitrary space-time point
with the quaternionic matrix
\beq
{\mathsf X}=
x^0
\left(
\barr{cc}
-1&0\\
0&-1\\
\earr
\right)
+
x^1
\left(
\barr{cc}
0&-\bi\\
\bi&0\\
\earr
\right)
+
x^2
\left(
\barr{cc}
0&-\bj\\
\bj&0\\
\earr
\right)
+
x^3
\left(
\barr{cc}
0&-\bk\\
\bk&0\\
\earr
\right),
\label{eqn:1-2}
\eeq
where the matrix with coefficient $x^0$
is changed
to minus the unit matrix for
convenience.
The apparent parallel between the spatial coordinates
and Hamilton's units,
and the clear separation between time and space
in (\ref{eqn:1-2}), as in (\ref{eqn:1-1}),
make it more natural
to consider ${\mathsf X}$ than $X$
as a position matrix 
and represent the Lorentz transformation
as a linear transformation,\footnote{This
possibility
was once considered by the present
author in Ref.~4), which, however, contains neither group theoretical
considerations nor the definition of
the quaternionic determinant.
Rather, it puts emphasis on the quaternionic Dirac theory
with less attention given to the intrusion
of $i$ into the theory.
A brief comment on the group properties
was made in Ref. 5).}
\beq
{\mathsf X}\to
{\mathsf X}'={\mathsf A}{\mathsf X}{\mathsf A}^\dag,
\label{eqn:1-3}
\eeq
provided that
det$\,{\mathsf X}$
is Lorentz invariant\footnote{It 
is seen in \S 4
that det$\,{\mathsf X}=[
(x^0)^2-(x^1)^2-(x^2)^2-(x^3)^2]^2$.}
and det$\,{\mathsf A}=1$.\footnote{We see in the next section
that we have det$\,{\mathsf A}^\dag=$det$\,{\mathsf A}$
for any $2\times 2$ quaternionic matrix.}
Unfortunately, however,
the matrix (\ref{eqn:1-2})
is not the most general
Hermitian matrix
in the field of real quaternions. (See the next section
for the definition of the Hermiticity
of a $2\times 2$ quaternionic matrix.)
It follows that
the unimodular matrix ${\mathsf A}$
in (\ref{eqn:1-3})
cannot be any element
of $SL(2,{\mathbb H})$.
We must have further conditions
in addition to unimodularity
to define the matrix ${\mathsf A}$,
but nevertheless retain the group properties
of the set of such ${\mathsf A}$.
If we can do this, we will have succeeded in 
representing
the Lorentz transformations
in terms of real quaternions only.\footnote{The possibility
of representing
the Lorentz transformations
in terms of real quaternions only
was first put forward by Dirac \cite{6)}, who employed the
projective method. Our work in Ref.~4)
was largely influenced by Dirac's 1945 paper.
Rotelli \cite{7)} independently found
a quaternionic representation
of the Lorentz group in a work proposing a quaternionic
Dirac equation.}
\\[2mm]
\ind
It is rather easy to find such conditions
by noting that to define
the most general $2\times 2$ 
Hermitian matrix
in the field of real quaternions,
we need two more matrices,
$\sigma_1$ and $\sigma_3$,
in addition to the 
matrices presented in (\ref{eqn:1-2}). 
This suggests that,
if these two matrices are invariant under the transformation by
${\mathsf A}$, i.e.
\beq
{\mathsf A}\sigma_1{\mathsf A}^\dag=\sigma_1,\;\;\;
{\mathsf A}\sigma_3{\mathsf A}^\dag=\sigma_3,
\label{eqn:1-4}
\eeq
then the transfomation
(\ref{eqn:1-3})
leads to the most general linear
transformation
for 4 real parameters, $x^i\;(i=0,1,2,3)$
with the constraint
det$\,{\mathsf X}'$=det$\,{\mathsf X}$.
We show in \S 4 that
the set of such ${\mathsf A}$
indeed does form a group,
denoted $Spin(2,{\mathbb H})$
in this paper,
which is 
locally isomorphic to the
restricted Lorentz group,
$L_+^\uparrow$.
\\[2mm]
\ind
Once we have
found the spinor group $Spin(2,{\mathbb H})$
over the field of real quaternions,
a next natural task is
to
reformluate
the Dirac theory
based on it.
The Dirac spinor
now consisting of
two-component
quaternions
undergoes {\it linear}
charge conjugation
transformation,
which means that
the Pauli-G\"ursey
$SU(2)$
group \cite{8)}
can be interpreted
as a right translation
of the
quaternionic Dirac spinor. 
\\[2mm]
\ind
The existence of the Pauli-G\"ursey $SU(2)$
group \footnote{Recall
that
the Pauli-G\"ursey $SU(2)$
group
is a symmetry group of the massless
Dirac Lagrangian
which, in addition, possesses the familiar
chiral symmetry.}
and the Hermiticity requirement
of the Dirac Lagrangian
constructed from
the two-component quaternionic
Dirac spinor
have the important
consequence
that
an additinoal imaginary unit
that commutes with
Hamilton's units
must be introduced
into the theory.
Taking this additional
imaginary
unit
to be the ordinary
imaginary unit $i$,
the intrusion
of $i$ into the theory
allows us to
make use of the Pauli
representation
(involving $i$)
of quaternions
and, moreover,
allows us to construct
the scheme
of complex quaternions
via diagonalization
of the matrix
(\ref{eqn:1-2})
and ${\mathsf A}$ of 
(\ref{eqn:1-4}).
It is shown in \S 6 that
the diagonalization
transforms the matrix
(\ref{eqn:1-2})
into the diagonal form
\beq
x^0
\left(
\barr{cc}
-1&0\\
0&-1\\
\earr
\right)
+
x^1
\left(
\barr{cc}
-i\bi&0\\
0&i\bi\\
\earr
\right)
+
x^2
\left(
\barr{cc}
-i\bj&0\\
0&i\bj\\
\earr
\right)
+
x^3
\left(
\barr{cc}
-i\bk&0\\
0&i\bk\\
\earr
\right).
\label{eqn:1-5}
\eeq
Consequently,
it is sufficient
to consider Hermitian quaternion\footnote{A Hermitian
quaternion
is a complex quaternion
which is invariant under the operations
$i\to -i$, in addition
to
$\bi,\bj,\bk\to
-\bi,-\bj,-\bk$,
respectively.}
\beq
{\mathsf x}=x^0+x^1(i\bi)
+x^2(i\bj)
+x^3(i\bk),
\label{eqn:1-6}
\eeq
which has the indefinite
norm squared
$(x^0)^2-(x^1)^2-(x^2)^2-(x^3)^2$. 
The matrix ${\mathsf A}$
is similarly diagonalized, simplifying
the transformation
(\ref{eqn:1-3}) with (\ref{eqn:1-4})
to
\beq
{\mathsf x}\to {\mathsf x}'=U{\mathsf x}U^\dag,
\label{eqn:1-7}
\eeq
where the squared norm of ${\mathsf x}'$
is eqaul to the
squared norm of ${\mathsf x}$, and
$U\in SL(1,{\mathbb H}^c)$, the
set of unit complex quaternions.
In this way, we naturally arrive at
the conventional approach
of employing complex quaternions
to represent the Lorentz transformations
and reformulate the Dirac theory
over ${\mathbb H}^c$,
the algebra of complex quaternions.\footnote{Our fomulation
suggests, therefore, that
the scheme of complex quaternions
used in relativity
is not to be assumed from the outset
but to be considered as a
natural mathematical method
inherited from the $Spin(2,{\mathbb H})$
Dirac theory.}
Considering
the isomorphism
$SL(1,{\mathbb H}^c)\cong SL(2,{\mathbb C})$,
we also find a relation
to the
spinor analysis (over the complex field).
Using the Pauli representation
of quaternions,
it turns out that
(\ref{eqn:1-6})
is identical to
(\ref{eqn:1-1}),
while (\ref{eqn:1-7})
is merely
the transformation,
$X\to X'=AXA^\dag$
for
$A\in SL(2,{\mathbb C})$.
(The details concerning these points are given in \S 6.)
\\[2mm]
\ind
This paper
is arranged
as follows.
In the next section,
we give 
a brief 
survey of
the basic
properties
of real quaternions
and of the
$2\times 2$ quaternionic matrix.
We then present
a proof of the
isomorphism
$SL(2,{\mathbb H})\cong{\overline{SO}}(5,1)$.
This isomorphism is well known;
the proof appearing
in \S 3 uses arguments similar to those in
spinor analysis.
This proof is used in
\S 4 to establish the
isomorphism
$Spin(2,{\mathbb H})/Z_2\cong L_+^\uparrow$.
On the basis of this isomorphism,
we reformulate
the Dirac theory
in \S 5,
where
the neccessity
of the additional imaginary unit
in the theory
is pointed out,
and
it is explicitly
shown that
the factors in the Dirac Lagrangian
are automatically
`anti-symmetrized',
as postulated in the proof of the $CPT$ theorem. \cite{9)}
This automatic `anti-symmetrization'
comes from the linearity
of the charge conjugation
transformation for the
quaternionic Dirac spinor.
In \S 6 we diagonalize the matrix ${\mathsf A}$
using a complex unitary matrix
and discuss the natural appearance
of complex quaternions
in the theory.
It should be remarked here
that,
in our approach,
the incorporation of $i$
into the theory
is not due to the geometry
of Minkowski space-time
but due entirely to the
consistency requirement in
the quaternionic Dirac theory
in the Lagrangian formalism.
In the final section
we 
summarize our results,
making some comments on
the application
of complex quaternions
to the Dirac theory
in general relativity.
Appendix A collects
some interesting
properties
of the matrix
${\mathsf A}$ in 
(\ref{eqn:1-4}), which were reported
in Ref. 4) without proof.
\section{Quaternions and $M(2,{\mathbb H})$}                            %
To define our notation, we briefly recall the basic properties of the
algebra of quaternions, ${\mathbb H}$. 
Any quaternion $q\in {\mathbb H}$ consists of four real numbers
and can be written in the form
\beq
q=q_0+\bi q_1+\bj q_2+\bk q_3
=q_0+q_1\bi+q_2\bj+q_3\bk,
\;\;\;q_i\in{\mathbb R},
\label{eqn:2-1}
\eeq
where Hamilton's imaginary units
$\bi, \bj$ and $\bk$ satisfy the equations
\beq
\bi^2=\bj^2=\bk^2=-1,\;\;\;
\bi\bj=\bk=-\bj\bi.
\hspace{1cm}
\mbox{(cyclic)}
\label{eqn:2-2}
\eeq
Hence, the algebra ${\mathbb H}$ is not
commutative and associative, as, for instance,
$(\bi\bj)\bk=\bi(\bj\bk)$.
The quaternionic conjugation ${\bar q}$ of $q$
is
defined by
\beqa
{\bar q}&=&q_0-\bi q_1-\bj q_2-\bk q_3,\nn\\[2mm]
\overline{pq}&=&{\bar q}{\bar p},\hspace{1cm}
p, q\in{\mathbb H}.
\label{eqn:2-3}
\eeqa
The real and imaginary parts of $q$ are given, 
respectively, by
\begin{eqnarray}
\mbox{Re}\,q\,=
\frac 12(q+{\bar q})
=\mbox{Re}\,{\bar q}\,,\hspace{1cm}
\mbox{Im}\,q\,=\frac 12(q-{\bar q})
=-\mbox{Im}\,{\bar q}\,.
\label{eqn:2-4}
\end{eqnarray}
Although quaternions are not commutative,
the real part of the product of two factors
is independent of the order of the factors:
\beq
\mbox{Re}[\,q\,p\,]=
\mbox{Re}[\,p\,q\,],\hspace{1cm}
p, q\in{\mathbb H}.
\label{eqn:2-5}
\eeq
This is generalized to the cyclic property of 
the operation Re[\,$\cdots$\,]
involving an arbitary number of factors.
The norm of a quaternion $q$,
\beq
N(\,q\,)=q{\bar q}={\bar q}q\equiv
|q|^2=q_0^2+q_1^2+q_2^2+q_3^2,
\label{eqn:2-6}
\eeq
is non-negative, vanishes iff $q=0$,
and possesses the composition property
\beq
N(\,pq\,)=N(\,p\,)N(\,q\,),\hspace{1cm}
p, q\in{\mathbb H}.
\label{eqn:2-7}
\eeq
This means that ${\mathbb H}$ is a division algebra.
The inverse of a non-zero quaternion $q$
is $q^{-1}={\bar q}/|q|^2$,
and we have $(pq)^{-1}=q^{-1}p^{-1}$.
\\[2mm]
\ind
Before going on, it is convenient to
adopt the following notation for the 
algebra ${\mathbb H}$:
\beq
1=e_0,\;\;\;
\bi=e_1,\;\;\;
\bj=e_2,\;\;\;
\bk=e_3.
\label{eqn:2-8}
\eeq
Then we have
\beq
{\mathbb H}=\{q=q_0e_0+q_1e_1+q_2e_2+q_3e_3
\equiv q_ie_i\;(i=0,1,2,3)
;\,
q_0, q_1, q_2, q_3\in{\mathbb R}\}.
\label{eqn:2-9}
\eeq
Note that $e_0$ is the unit element of the algebra ${\mathbb H}$; 
that is, we have $e_0e_0=e_0$ and $e_0\be=\be e_0$, 
with $\be=(e_1,e_2,e_3)$,
and (\ref{eqn:2-2}) 
becomes
\beq
e_ae_b=-\delta_{ab}e_0+\epsilon_{abc}e_c,\hspace{1cm}
a, b, c=1, 2, 3,
\label{eqn:2-10}
\eeq
where $\epsilon_{abc}$ is the 3-dimensional
Levi-Civita symbol with 
$\epsilon_{123}=1$.
\\[2mm]
\ind
Any quaternion $p$ can also be written as
\beq
p=e_0(p_0+p_3e_3)+e_2(p_2+p_1e_3)
\equiv e_0\cP_1+e_2\cP_2,
\label{eqn:2-11}
\eeq
with $\cP_1=p_0+p_3e_3$ and $\cP_2=p_2+p_1e_3$.
This is called the symplectic representation.
If we write
\beqa
qe_0&=&e_0(q_0+q_3e_3)+e_2(q_2+q_1e_3),
\nn\\[2mm]
qe_2&=&e_0(-q_2+q_1e_3)+e_2(q_0-q_3e_3),\hspace{1cm}q\in{\mathbb H},
\label{eqn:2-12}
\eeqa
we easily find
\beq
qp=(e_0, e_2)\left(
             \barr{cc}
             q_0+q_3e_3&-q_2+q_1e_3\\
             q_2+q_1e_3&q_0-q_3e_3\\
             \earr
             \right)
\left(
\barr{l}
\cP_1\\
\cP_2\\
\earr
\right).
\label{eqn:2-13}
\eeq
This implies that the mapping
$p\to qp$ with $q\in{\mathbb H}$, an endomorphism of ${\mathbb H}$,
induces the linear transformation
\beq
\left(
\barr{l}
\cP_1\\
\cP_2\\
\earr
\right)
\to
\left(
\barr{l}
\cP\,'_1\\
\cP\,'_2\\
\earr
\right)
=
\left(
             \barr{cc}
             q_0+q_3e_3&-q_2+q_1e_3\\
             q_2+q_1e_3&q_0-q_3e_3\\
             \earr
             \right)\left(
\barr{l}
\cP_1\\
\cP_2\\
\earr
\right).
\label{eqn:2-14}
\eeq
The transformation matrix in (\ref{eqn:2-14})
is a representation of $q\in{\mathbb H}$,
since $e_3^2=-1$ and
$q_ie_3=e_3q_i\;(i=0,1,2,3)$. 
Now, it is important to recognize that
the transformation matrix in (\ref{eqn:2-14}) is complex,
because the field ${\mathbb C}(1, e_3)$ spanned by
1 and $e_3$
is isomorphic to
the complex field ${\mathbb C}$.
However, there is an ambiguity in the identification of
$e_3$ to establish the isomorphism.
Specifically, it could be either $\sqrt{-1}$
or $-\sqrt{-1}$, and these two identifications 
lead to
two representations
\beqa
\!\!\!\!\!\!\!\!\!\!
\rho(q)
&=&
\left(
             \barr{cc}
             q_0+q_3e_3&-q_2+q_1e_3\\
             q_2+q_1e_3&q_0-q_3e_3\\
             \earr
             \right)\Biggr|_{e_3=-\sqrt{-1}}
=
\left(
             \barr{cc}
             q_0-iq_3&-q_2-iq_1\\
             q_2-iq_1&q_0+iq_3\\
             \earr
             \right),\\[3mm]
\!\!\!\!\!\!\!\!\!\!
{\bar\rho}(q)
&=&
\left(
             \barr{cc}
             q_0+q_3e_3&-q_2+q_1e_3\\
             q_2+q_1e_3&q_0-q_3e_3\\
             \earr
             \right)\Biggr|_{e_3=\sqrt{-1}}
=
\left(
             \barr{cc}
             q_0+iq_3&-q_2+iq_1\\
             q_2+iq_1&q_0-iq_3\\
             \earr
             \right),
\label{eqn:2-15}
\eeqa
where we use the conventional notation $i\equiv\sqrt{-1}$.
We then have \footnote{The similar relation
$\mbox{Re}\,q=\frac 12\mbox{tr}\,{\bar\rho}(q)$
will not be used below.}
\beq
\mbox{Re}\,q=\frac 12\mbox{tr}\,\rho(q).
\label{eqn:2-17}
\eeq
These two representations are equivalent,
because $\rho(q)$ and ${\bar\rho}(q)$ satisfy
\beq
\omega {\bar\rho}(q)\omega^{-1}
=\rho(q),\;\;\;
\omega=\left(
\barr{cc}
0&-1\\
1&0\\
\earr
\right).
\label{eqn:2-18}
\eeq
This is analogous to $A\sim A^*$ if $A\in SU(2)$.\footnote{For a real 
quaternion $q$, we simply have ${\bar\rho}(q)=\rho^*(q)$. Although
(\ref{eqn:2-18}) is still valid for a complex quaternion $q$,
the equality ${\bar\rho}(q)=\rho^*(q)$
no longer holds for a complex quaternion.
This fact has a non-trivial implication, as we see in \S 6.
This is the reason that we express the two representations
with different notation.}
For the basis (\ref{eqn:2-8}), we find
\begin{eqnarray}
\rho(e_0)&=&\sigma_0=\left(
\barr{cc}
1&0\\
0&1\\
\earr
\right),\;\;\;
\rho(e_1)=-i\sigma_1=\left(
\barr{cc}
0&-i\\
-i&0\\
\earr
\right),\nn\\[2mm]
\rho(e_2)&=&-i\sigma_2=\left(
\barr{cc}
0&-1\\
1&0\\
\earr
\right),\;\;\;
\rho(e_3)=-i\sigma_3=\left(
\barr{cc}
-i&0\\
0&i\\
\earr
\right),
\label{eqn:2-19}
\end{eqnarray}
with $\sigma_i=(\sigma_0, \sigma_1, \sigma_2, \sigma_3)$
being the Pauli matrix (with the unit matrix $\sigma_0$ included).
Accordingly, we call the representation
$\rho(q)$ of (2$\cdot$15)
the Pauli representation of a quaternion $q$.
Note that it
involves $i$.
Similarly, we have
\begin{eqnarray}
{\bar\rho}(e_0)&=&{\tilde\sigma}_0^*=\left(
\barr{cc}
1&0\\
0&1\\
\earr
\right),\;\;\;
{\bar\rho}(e_1)=-i{\tilde\sigma}_1^*=\left(
\barr{cc}
0&i\\
i&0\\
\earr
\right),\nn\\[2mm]
{\bar\rho}(e_2)&=&-i{\tilde\sigma}_2^*=\left(
\barr{cc}
0&-1\\
1&0\\
\earr
\right),\;\;\;
{\bar\rho}(e_3)=-i{\tilde\sigma}_3^*=\left(
\barr{cc}
i&0\\
0&-i\\
\earr
\right),
\label{eqn:2-20}
\end{eqnarray}
with ${\tilde\sigma}_i=(\sigma_0, -\sigma_1, -\sigma_2, -\sigma_3)$.
The matrix ${\tilde\sigma}_i$
satisfies the identity
${\tilde\sigma}_i=\omega\sigma_i^T\omega^{-1}$.
\\[2mm]
\ind
There is another representation given by the
4-dimensional complex matrix
\begin{eqnarray}
\omega(e_0)&=&\Omega_0
\equiv\left(
\barr{cccc}
1&0&0&0\\
0&1&0&0\\
0&0&1&0\\
0&0&0&1\\
\earr
\right),\;
\omega(e_1)=-i\Omega_1
\equiv
\left(
\barr{cccc}
0&-i&0&0\\
-i&0&0&0\\
0&0&0&-1\\
0&0&1&0\\
\earr
\right),\nn\\[2mm]
\omega(e_2)&=&-i\Omega_2
\equiv
\left(
\barr{cccc}
0&0&-i&0\\
0&0&0&1\\
-i&0&0&0\\
0&-1&0&0\\
\earr
\right),\;
\omega(e_3)=-i\Omega_3
\equiv
\left(
\barr{cccc}
0&0&0&-i\\
0&0&-1&0\\
0&1&0&0\\
-i&0&0&0\\
\earr
\right).
\label{eqn:2-21}
\end{eqnarray}
This representation is not irreducible, but it is a
direct sum of the above representations:
\beq
V\omega(q)V^{-1}
=\left(
\barr{cc}
\rho(q)&0\\
0&{\bar\rho}(q)\\
\earr
\right),\;\;\;
V=
\left(
\barr{cccc}
1&0&0&1\\
0&1&i&0\\
1&0&0&-1\\
0&-1&i&0\\
\earr
\right).
\label{eqn:2-22}
\eeq
This fact is used in \S 6.
\\[2mm]
\ind
We also consider the $2\times 2$ quaternionic matrix
\beq
{\mathbf A}=
\left(
\barr{cc}
a&b\\
c&d\\
\earr
\right),\;\;\;
a, b, c, d\in{\mathbb H}.
\label{eqn:2-23}
\eeq
The set of ${\mathbf A}$ is
called $M(2, {\mathbb H})$.
Its quaternionic conjugation, ${\bar{\mathbf A}}$,
is obtained by
taking the quaternionic conjugation
of all elements, and thus
\beq
\mbox{Re}\,{\mathbf A}=
\frac 12({\mathbf A}+{\bar{\mathbf A}})
=\left(
\barr{cc}
\mbox{Re}\,a&\mbox{Re}\,b\\
\mbox{Re}\,c&\mbox{Re}\,d\\
\earr
\right).
\label{eqn:2-24}
\eeq
We also define
\beq
\mbox{Tr}\,{\mathbf A}
=\mbox{tr (Re}\,
{\mathbf A})
=\mbox{Re (tr}\,
{\mathbf A}),
\label{eqn:2-25}
\eeq
where tr representss the matrix trace.
Hermitian conjugation of ${\mathbf A}$
is defined by
\beq
{\mathbf A}^\dag=
{\bar{\mathbf A}}^T=
\left(
\barr{cc}
{\bar a}&{\bar c}\\
{\bar b}&{\bar d}\\
\earr
\right).
\label{eqn:2-26}
\eeq
If ${\mathbf A}={\mathbf A}^\dag$, ${\mathbf A}$ 
is said to be Hermitian.
The set of Hermitian matrices is denoted by
$H(2,{\mathbb H})=\{{\mathbf A}\in M(2,{\mathbb H});
\;{\mathbf A}^\dag={\mathbf A}\}$.
Any ${\mathbf A}\in H(2,{\mathbb H})$ has real diagonal elements, 
and thus
Tr${\mathbf A}=$tr${\mathbf A}$.
From the cyclic property of the operation Re$\,[\cdots]$,
we have 
\beq
\mbox{Tr}\,({\mathbf A}{\mathbf B})
=\mbox{Tr}\,({\mathbf B}{\mathbf A}),\;\;\;
{\mathbf A},\;
{\mathbf B}\in M(2,{\mathbb H}).
\label{eqn:2-27}
\eeq
In contrast to the case of a tranposed matrix,\footnote{Due to
the noncommutativity of quaternions, in general we have
$({\mathbf A}_1{\mathbf A}_2)^T
\ne {\mathbf A}_2^T{\mathbf A}_1^T$.}
we have
\beq
({\mathbf A}_1{\mathbf A}_2)^\dag
={\mathbf A}_2^\dag{\mathbf A}_1^\dag.
\label{eqn:2-28}
\eeq
The quaternionic determinant
is defined by
\beq
\mbox{det}\,{\mathbf A}
=\mbox{det}\,\rho({\mathbf A})
\equiv
\mbox{det}\,\left(
\barr{cc}
\rho(a)&\rho(b)\\
\rho(c)&\rho(d)\\
\earr
\right),
\label{eqn:2-29}
\eeq
where $\rho$ is the representation ({2$\cdot$15).
Because $\rho$ satisfes
$\rho(a_1a_2+b_1c_2)=\rho(a_1a_2)+\rho(b_1c_2)
=\rho(a_1)\rho(a_2)+\rho(b_1)\rho(c_2)$, 
the following composition relation holds:
\beq
\mbox{det}\,({\mathbf A}_1{\mathbf A}_2)
=(\mbox{det}\,{\mathbf A}_1)(\mbox{det}\,{\mathbf A}_2).
\label{eqn:2-30}
\eeq
By factorizing the matrix (\ref{eqn:2-23})
and using the composition property (\ref{eqn:2-30}),
we find
\beq
\mbox{det}\,\left(
\barr{cc}
a&b\\
c&d\\
\earr
\right)
=
\mbox{det}\,\left(
\barr{cc}
a-bd^{-1}c&bd^{-1}\\
0&1\\
\earr
\right)
\mbox{det}\,\left(
\barr{cc}
1&0\\
c&d\\
\earr
\right).
\label{eqn:2-31}
\eeq
We then make use of the definition
(\ref{eqn:2-29})
to obtain the result
\beq
\mbox{det}\,{\mathbf A}
=
\mbox{det}\,\left(
\barr{cc}
a&b\\
c&d\\
\earr
\right)
=
|a-bd^{-1}c|^2|d|^2,
\label{eqn:2-32}
\eeq
since
det$\,\left(
\barr{cc}
a&b\\
0&1\\
\earr
\right)
=|a|^2$ and
det$\,\left(
\barr{cc}
1&b\\
0&d\\
\earr
\right)
=|d|^2$.
This relation was quoted by G\"ursey \cite{10)}
without proof and is called the Study determinant.
\footnote{The author
is grateful to Professor T. Suzuki for informing him
of Ref.~11).}
\\[2mm]
\ind
There are several equivalent expressions for the
quaternionic determinant (\ref{eqn:2-32}).
Among them, we present two here:
\beqa
\mbox{det}\,\left(
\barr{cc}
a&b\\
c&d\\
\earr
\right)
&=&
|a|^2|d|^2+|b|^2|c|^2-2\mbox{Re}\,[a{\bar c}d{\bar b}]\nn\\[2mm]
&=&
(|a|^2-|b|^2)(|d|^2-|c|^2)
+|a{\bar c}-b{\bar d}|^2.
\label{eqn:2-33}
\eeqa
These are called the first (\ref{eqn:2-32}), 
second and third alternatives (\ref{eqn:2-33})
of the quaternionic determinant.
Using the second alternative,
we obtain
\beq
\mbox{det}\,{\mathbf A}
=\mbox{det}\,{\mathbf A}^\dag.
\label{eqn:2-34}
\eeq
The inverse of the matrix ${\mathbf A}$
is given by
\beq
{\mathbf A}^{-1}
=\frac 1{\mbox{det}\,{\mathbf A}}
\left(
\barr{cc}
|d|^2{\bar a}-{\bar c}d{\bar b}&
|b|^2{\bar c}-{\bar a}b{\bar d}\\
|c|^2{\bar b}-{\bar d}c{\bar a}&
|a|^2{\bar d}-{\bar b}a{\bar c}\\
\earr
\right),\hspace{0.5cm}
\mbox{det}\,{\mathbf A}\ne 0,
\label{eqn:2-35}
\eeq
where we repeatedly used the second alternative
of the quaternionic determinant.
The inverse matrix satisfies the usual relation
\beq
({\mathbf A}_1{\mathbf A}_2)^{-1}
={\mathbf A}_2^{-1}{\mathbf A}_1^{-1}.
\label{eqn:2-36}
\eeq
\section{$SL(2, {\mathbb H})$ as the spinor group of the conformal
group in ${\mathbb R}^4$}                                          %
\ind
It is well known that
the group
\beq
SL(2, {\mathbb H})=
\Big\{{\mathbf A}\in M(2, {\mathbb H});\;\mbox{det}\,
{\mathbf A}=1\Big\}
\label{eqn:3-1}
\eeq
is the universal covering group of
the conformal group
in Euclidean 4-space ${\mathbb R}^4$:
\beq
SL(2, {\mathbb H})\cong\overline{SO}(5, 1).
\label{eqn:3-2}
\eeq
For later convenience
in this section we present
an explicit proof of (\ref{eqn:3-2})
in a manner similar to the
proof of $SL(2, {\mathbb C})\cong\overline{SO}(3, 1)$.
\\[2mm]
\ind
Any $2\times 2$ quaternionic Hermitian matrix ${\cal X}
={\cal X}^\dag$
has six independent real numbers,
two real diagonal elements
and one off-digaonal element, a
quaternion with four real numbers.
Hence, it can be decomposed into
a sum of six independent Hermitian matrices
with real coefficients as 
\beq
{\cal X}=X^0\Gamma_0
+X^1\Gamma_1+X^2\Gamma_2+
X^3\Gamma_3+X^4\Gamma_4+
X^5\Gamma_5\equiv X^I\Gamma_I,\;\;\;
X^I\;(I=0,1,\cdots, 5)\in {\mathbb R},
\label{eqn:3-3}
\eeq
where
\beqa
\Gamma_0&=&
\left(
\barr{cc}
-1&0\\
0&-1\\
\earr
\right),\;\;\;
\Gamma_1=
\left(
\barr{cc}
0&-\bi\\
\bi&0\\
\earr
\right),\;\;\;
\Gamma_2=
\left(
\barr{cc}
0&-\bj\\
\bj&0\\
\earr
\right),\nn\\[3mm]
\Gamma_3&=&
\left(
\barr{cc}
0&-\bk\\
\bk&0\\
\earr
\right),\;\;\;
\Gamma_4=
\left(
\barr{cc}
0&1\\
1&0\\
\earr
\right),\;\;\;
\Gamma_5=
\left(
\barr{cc}
1&0\\
0&-1\\
\earr
\right).
\label{eqn:3-4}
\eeqa
We use the notation
$\sigma_0=-\Gamma_0,
\sigma_1=\Gamma_4$
and $\sigma_3=\Gamma_5$,
while
the second Pauli matrix $\sigma_2$
is generalized to $\Gamma_{1,2,3}$
by replacing $i$ with Hamilton's units, $\bi, \bj, \bk$.
Thus, the six $\Gamma_I\;(I=0,1,\cdots,5)$ matrices are simply
the Pauli matrices in $D=6$ space-time.
It is easy to prove the following:
\beqa
\mbox{det}\,{\cal X}
&=&(-X^2)^2,\nn\\[2mm]
X^2&=&-(X^0)^2+(X^1)^2+(X^2)^2
+(X^3)^2+(X^4)^2+(X^5)^2
\equiv \eta_{IJ}X^IX^J,\nn\\[2mm]
\eta&=&(\eta_{IJ})=
\mbox{diag}\,(-1, +1, +1, +1, +1, +1).
\label{eqn:3-5}
\eeqa
We use the metric $\eta_{IJ}=\eta^{IJ}$
to lower or raise the indices $I, J, K\cdots$,
which run from 0 to 5.
The most general transformation
from a Hermitian matrix to
another Hermitian matrix without changing the determinant
is given by
\beq
{\cal X}\to{\cal X}'={\mathbf A}{\cal X}{\mathbf A}^\dag,\;\;\;
{\mathbf A}\in SL(2, {\mathbb H}).
\label{eqn:3-6}
\eeq
If we write
\beq
{X'}^I=\Lambda^I_{\;\;J}X^J,\;\;\;
\eta_{IJ}\Lambda^I_{\;\;K}
\Lambda^J_{\;\;L}=\eta_{KL},
\label{eqn:3-7}
\eeq
we immediately obtain
\beq
\Lambda^I_{\;\;J}\Gamma_I
={\mathbf A}\Gamma_J{\mathbf A}^\dag,\;\;\;
{\mathbf A}\in SL(2, {\mathbb H}),
\label{eqn:3-8}
\eeq
where $\Lambda^I_{\;\;J}$ is now a continuous function
of ${\mathbf A}: \Lambda^I_{\;\;J}=\Lambda^I_{\;\;J}({\mathbf A})$.
\\[2mm]
\ind
If we invert the matrix ${\cal X}$,
provided that $X^2\ne 0$,
we obtain 
\beq
{\cal X}^{-1}
=\frac 1{-X^2}{\tilde {\cal X}},\;\;\;
{\tilde {\cal X}}\equiv X^I{\tilde\Gamma}_I,
\label{eqn:3-9}
\eeq
where
\beq
{\tilde \Gamma}_0=\Gamma_0,\;\;\;
{\tilde \Gamma}_I=-\Gamma_I.\;\;\;(I=1,2,\cdots, 5)
\label{eqn:3-10}
\eeq
Consequently, we have, since $X^2={X'}^2$, 
\beq
\Lambda^I_{\;\;J}{\tilde\Gamma}_I
={\mathbf A}^{\dag\,-1}{\tilde\Gamma}_J{\mathbf A}^{-1},\;\;\;
{\mathbf A}\in SL(2, {\mathbb H}).
\label{eqn:3-11}
\eeq
The matrices $\{\Gamma_I, {\tilde\Gamma}_J\}_{I,J=0,1,\cdots, 5}$
satisfy the relations
\beq
\Gamma_I{\tilde\Gamma}_J
+\Gamma_J{\tilde\Gamma}_I=-2\eta_{IJ}1_2,\;\;\;
{\tilde\Gamma}_I\Gamma_J
+{\tilde\Gamma}_J\Gamma_I=-2\eta_{IJ}1_2,
\label{eqn:3-12}
\eeq
where $1_2=\left(
           \barr{cc}
           1&0\\
           0&1\\
           \earr
           \right)$
is the unit matrix in $SL(2,{\mathbb H})$.
Sandwiching the first of these equations
between ${\mathbf A}$ (from left)
and ${\mathbf A}^{-1}$ (from right)
we find from (\ref{eqn:3-8}) and (\ref{eqn:3-11})
that
$\eta_{IJ}\Lambda^I_{\;\;K}
\Lambda^J_{\;\;L}=\eta_{KL}$,
which implies $X^2={X'}^2$.
Hence $\big(\Lambda^I_{\;\;J}({\mathbf A})
\big)\in O(5, 1)$.
We next have to show the following:
\\[2mm]
\hspace{1cm}(i) $\Lambda^0_{\;\;0}\ge 1$;\\[2mm]
\hspace{1cm}(ii) det$\,(\Lambda^I_{\;\;J})=1$;\\[2mm]
\hspace{1cm}(iii) $\Lambda^I_{\;\;J}({\mathbf A}_1)
      \Lambda^J_{\;\;K}({\mathbf A}_2)=\Lambda^I_{\;\;K}
      ({\mathbf A}_1{\mathbf A}_2)$;\\[2mm]
\hspace{1cm}(iv) $\Lambda^I_{\;\;J}({\mathbf A})
=\Lambda^I_{\;\;J}(-{\mathbf A})$.\\[2mm]
\ind
To prove the above results, we first note that, using
Tr$\,[\Gamma_I{\tilde\Gamma}_J]=-2\eta_{IJ}$, 
from the definition
(\ref{eqn:3-4}) and (\ref{eqn:3-10}),
(\ref{eqn:3-8}) leads to
\begin{equation}
\Lambda^I_{\;\;J}
=-\frac 12\mbox{Tr}\,[{\mathbf A}
\Gamma_J{\mathbf A}^\dag
{\tilde\Gamma}^I].
\label{eqn:3-13}
\end{equation}
First, we have $\Lambda^0_{\;\;0}
=\frac 12\mbox{Tr}\,[{\mathbf A}{\mathbf A}^\dag]
>0$, and hence $\Lambda^0_{\;\;0}\ge 1$ from $\eta_{IJ}
\Lambda^J_{\;\;K}\Lambda^J_{\;\;L}=\eta_{KL}$.
This is the result (i).
To prove the result (ii), we make use of the formula 
\beq
\mbox{Tr}\,[\Gamma_I{\tilde\Gamma}_J\Gamma_K{\tilde\Gamma}_L
\Gamma_M{\tilde\Gamma}_N]
=15 \;\mbox{terms containing}\;\eta_{IJ}
+2\epsilon_{IJKLMN},
\label{eqn:3-14}
\eeq
where $\epsilon_{IJKLMN}$ is the 6-dimensional
Levi-Civita
symbol with $\epsilon_{012345}=-1$.
Now, using (\ref{eqn:3-8}) and (\ref{eqn:3-11}),
we compute 
\beqa
&&(\Lambda^I_{\;\;0}\Gamma_I)
(\Lambda^J_{\;\;1}{\tilde\Gamma}_J)(\Lambda^K_{\;\;2}
\Gamma_K)(\Lambda^L_{\;\;3}{\tilde\Gamma}_L)
(\Lambda^M_{\;\;4}\Gamma_M)(\Lambda^N_{\;\;5}{\tilde\Gamma}_N)\nn\\[2mm]
&&=
({\mathbf A}\Gamma_0{\mathbf A}^\dag)
({\mathbf A}^{\dag\,-1}{\tilde\Gamma}_1{\mathbf A}^{-1})
({\mathbf A}\Gamma_2{\mathbf A}^\dag)
({\mathbf A}^{\dag\,-1}{\tilde\Gamma}_3{\mathbf A}^{-1})
({\mathbf A}\Gamma_4{\mathbf A}^\dag)
({\mathbf A}^{\dag\,-1}{\tilde\Gamma}_5{\mathbf A}^{-1})\nn\\[2mm]
&&
={\mathbf A}(\Gamma_0{\tilde\Gamma}_1\Gamma_2
{\tilde\Gamma}_3\Gamma_4{\tilde\Gamma}_5){\mathbf A}^{-1}.
\label{eqn:3-15}
\eeqa
Taking the trace Tr of this equation,
no term containing an $\eta_{IJ}$ in
(\ref{eqn:3-14}) contributes to
the left-hand side, as can be shown by making repeated use of
$\eta_{IJ}
\Lambda^I_{\;\;K}\Lambda^J_{\;\;L}=\eta_{KL}$,
while the matrix ${\mathbf A}$ disappears from the right-hand side.
We are then left with
\beqa
&&
-\epsilon_{IJKLMN}
(\Lambda^I_{\;\;0}
\Lambda^J_{\;\;1}
\Lambda^K_{\;\;2}\Lambda^L_{\;\;3}
\Lambda^M_{\;\;4}\Lambda^N_{\;\;5})
=+1,
\label{eqn:3-16}
\eeqa
which implies det$\,(\Lambda^I_{\;\;J})=1$, the result (ii). 
To prove (iii), we write
\beq
\Lambda^I_{\;\;J}({\mathbf A}_1)
\Lambda^J_{\;\;K}({\mathbf A}_2)
=\big(-\frac 12\big)^2
\Biggl(\mbox{Tr}\,[\alpha^I\Gamma_J]\Biggr)
\Biggl(\mbox{Tr}\,[\beta_K{\tilde\Gamma}^J]\Biggr),
\label{eqn:3-17}
\eeq
where $\alpha^I={\mathbf A}^\dag_1{\tilde\Gamma}^I{\mathbf A}_1$
and
$\beta_K={\mathbf A}_2\Gamma_K{\mathbf A}_2^\dag$.
Using the fact that
both $\alpha^I$ and $\beta_K$ are Hermitian,
it is straightforward to show that
\beqa
&&\big(-\frac 12\big)^2
\Biggl(\mbox{Tr}\,[\alpha^I\Gamma_J]\Biggr)
\Biggl(\mbox{Tr}\,[\beta_K{\tilde\Gamma}^J]\Biggr)\nn\\[2mm]
&&=-\frac 12
\mbox{Tr}\,[\alpha^I\beta_K]
=-\frac 12
\mbox{Tr}\,[({\mathbf A}_1{\mathbf A}_2)^\dag
{\tilde\Gamma}^I({\mathbf A}_1{\mathbf A}_2)\Gamma_K]
=\Lambda^I_{\;\;K}({\mathbf A}_1{\mathbf A}_2),
\label{eqn:3-18}
\eeqa
which proves (iii). The property (iv) is obvious.
Consequently,
we have proved the isomorphism (\ref{eqn:3-2})
in a manner similar to that used in the spinor calculus.
\section{The restricted Lorentz group as a subgroup 
         of $SL(2, {\mathbb H})$}  
The group $SL(2, {\mathbb H})$ is wider than the Lorentz group;
$SL(2, {\mathbb H})$ has 15 real parameters, $\omega_{IJ}=
-\omega_{JI}\;(I,J=0,1,2,3,4,5)$, while
the Lorentz group possesses only 6 real parameters,
$\omega_{ij}=
-\omega_{ji}\;(i,j=0,1,2,3)$.
The spinor algebraic proof of the
isomorphism (\ref{eqn:3-2})
has a great advantage concerning the restriction
of $SL(2, {\mathbb H})$ to the Lorentz group. 
In this section,
we see how this restriction
is made.
\\[2mm]
\ind
The Lorentz transformations
are obtained from the conformal transformations
(\ref{eqn:3-7}) in ${\mathbb R}^4$
by stipulating
${X'}^4=\lambda X^4$ and ${X'}^5=\mu X^5$
for $\lambda, \mu\in{\mathbb R}$, which require that the planes
$X^4=X^5=0$ remain invariant.
That is,
we take the matrix elements
$\Lambda^I_{\;\;J}$ simply as follows:
\beq
\Lambda^I_{\;\;J}
=
\left\{
\barr{l}
\Lambda^i_{\;\;j}\;\;\;\mbox{for}\;\;\;I=i=0,1,2,3;J=j=0,1,2,3,\\
\lambda\;\;\;\;\;\;\mbox{for}\;\;\;I=J=4,\\
\mu\;\;\;\;\;\;\mbox{for}\;\;\;I=J=5,\\
0\;\;\;\;\;\;\mbox{otherwise}.\\
\earr
\right.
\label{eqn:4-1}
\eeq
First,
we obtain the result that
(\ref{eqn:3-8})
leads to the equations
\beq
{\mathbf A}\sigma_1{\mathbf A}^\dag=\lambda\sigma_1,\;\;\;
{\mathbf A}\sigma_3{\mathbf A}^\dag=\mu\sigma_3,
\label{eqn:4-2}
\eeq
where we regard
the Pauli matrices $\sigma_1=\Gamma_4$
and $\sigma_3=\Gamma_5$ as elements of $M(2,{\mathbb H})$.
Since det$\,(\Lambda^I_{\;\;J})=1$,
we have
\beq
(\lambda\mu)\mbox{det}\,\Lambda=1.
\label{eqn:4-3}
\eeq
Here and hereafter we define
$\Lambda\equiv(\Lambda^i_{\;\;j})$.
Taking the quaternionic determinant of (\ref{eqn:4-2}),
we must have
$\lambda^4=\mu^4=1$. 
Moreover, if ${\mathbf A}_1$ and ${\mathbf A}_2$
satisfy (\ref{eqn:4-2}), 
the
product ${\mathbf A}_1{\mathbf A}_2$ also satisfies
the same equations, provided that $\lambda^2=\lambda$ and
$\mu^2=\mu$.
Hence, $\lambda^2=\mu^2=1$.
There are, therefore, four cases,
$(\lambda, \mu)=(+1,+1), (+1,-1),
(-1,+1), (-1,-1)$,
which correspond,
respectively, to
det$\,\Lambda=1, -1, -1,+1$. 
Note that $\Lambda^0_{\;\;0}\ge 1$
is still valid in all cases.
Suppose that we are given the matrix ${\mathbf A}$
for given $(\lambda, \mu)$. Then the cases $(\lambda, -\mu),
(-\lambda, \mu), (-\lambda, -\mu)$ correspond to the matrices,
${\mathbf A}\sigma_1, {\mathbf A}\sigma_3, 
{\mathbf A}\sigma_1\sigma_3={\mathbf A}\omega$,
respectively.
However, only the case
$\lambda=\mu=1$ is connected to the
identity $(\Lambda^I_{\;\;J})=(\eta^I_{\;\;J})$.
\\[2mm]
\ind
Let us therefore consider only the connected component
$\lambda=\mu=+1$ in what follows:
\beq
{\mathbf A}\sigma_1{\mathbf A}^\dag=\sigma_1,\;\;\;
{\mathbf A}\sigma_3{\mathbf A}^\dag=\sigma_3.
\label{eqn:4-4}
\eeq
This is the condition
noted in (\ref{eqn:1-4}).
Using $\sigma_1=\omega\sigma_3$, we have ${\mathbf A}
\omega\sigma_3{\mathbf A}^\dag=\omega\sigma_3$ from the first equation.
Combining this with the second equation,
we have
${\mathbf A}\omega{\mathbf A}^{-1}=\omega$.
That is, ${\mathbf A}$ commutes with $\omega$.
Therefore, ${\mathbf A}=Q1_2+P\omega$, for
$Q, P\in{\mathbb H}$.
For this ${\mathbf A}$ we have
${\mathbf A}(\sigma_{1,3}{\mathbf A}^\dag)
={\mathbf A}({\bar{\mathbf A}}\sigma_{1,3})=\sigma_{1,3}$,
and hence ${\mathbf A}{\bar{\mathbf A}}=1_2$ which also leads to
${\bar{\mathbf A}}{\mathbf A}=1_2$.
This gives rise to the relations
$|Q|^2-|P|^2=1$ and Re$\,[Q{\bar P}]=0$.
Consequently, (\ref{eqn:4-4}) is satisfied 
by the matrix
\beqa
{\mathbf A}
&=&\left(
\barr{cc}
Q&-P\\
P&Q\\
\earr
\right)\equiv{\mathsf A},\;\;\;
{\mathsf A}^{-1}={\bar{\mathsf A}},
\nn\\[2mm]
&&|Q|^2-|P|^2=1,\;\;\;
\mbox{Re}\,[{\bar Q}P]=0,
\;\;\;Q, P\in{\mathbb H}.
\label{eqn:4-5}
\eeqa
The two conditions on the matrix elements
imply det$\,{\mathsf A}=1$
from the third alternative of the 
quaternionic determinant.
The set of matrices (\ref{eqn:4-5}) 
forms a 6-parameter group.\footnote{This group is denoted $G$
in Ref. 5).}
We call it $Spin(2,{\mathbb H})$\footnote{The 
spinor group consisting of
$2\times 2$ quaternionic matrices (\ref{eqn:4-5})
defined here is non-compact, in contrast to
the compact spinor group
$Spin(n)\;(n\ge 3)$, which may be written $Spin(n,{\mathbb R})$.}
in this paper.
It is a subgroup of $SL(2, {\mathbb H})$, because,
for any ${\mathsf A}_1, {\mathsf A}_2 \in Spin(2,{\mathbb H})$,
we have
\beq
{\mathsf A}_1{\mathsf A}_2^{-1}
={\mathsf A}_1{\bar {\mathsf A}}_2
=\left(
\barr{cc}
Q_1&-P_1\\
P_1&Q_1\\
\earr
\right)
\left(
\barr{cc}
{\bar Q}_2&-{\bar P}_2\\
{\bar P}_2&{\bar Q}_2\\
\earr
\right)
=\left(
\barr{cc}
Q_3&-P_3\\
P_3&Q_3\\
\earr
\right)\in Spin(2,{\mathbb H}),
\label{eqn:4-6}
\eeq
since $(Q_3=Q_1{\bar Q}_2-P_1{\bar P}_2, 
P_3=Q_1{\bar P}_2+P_1{\bar Q}_2)$ satisfies 
the two equations in (\ref{eqn:4-5})
if $(Q_i, P_i)\;\;(i=1,2)$ satisfy the same
equations.
In what follows, ${\mathsf A}$ always represents the matrix (\ref{eqn:4-5})
belonging to $Spin(2,{\mathbb H})$.
Some properties of this matrix
are summarized in Appendix A.
\\[2mm]
\ind
Second, we obtain
the following relationship between $Spin(2,{\mathbb H})$ and
the restricted Lorentz group in analogy to the spinor analysis.
We have already encountered the necessary formulae
in the previous section.
Disregading the coordinates $X^4$ and $X^5$
in the previous section,
we simply write $X^i\equiv x^i\;(i=0,1,2,3)$
and
\beq
{\mathsf X}=x^i\Gamma_i,
\label{eqn:4-7}
\eeq
where $\Gamma_i\;(i=0,1,2,3)$ are given by (\ref{eqn:3-4}).
The position matrix (\ref{eqn:4-7})
is identical to (\ref{eqn:1-2}).
Its determinant is, from (\ref{eqn:3-5}),
\beqa
\mbox{det}\,{\mathsf X}
=(-x^2)^2,\;\;\;
x^2=\eta_{ij}x^ix^j,\;\;\;
\eta=(\eta_{ij})=
\mbox{diag}\,(-1, +1, +1, +1).
\label{eqn:4-8}
\eeqa
We use the metric $\eta_{ij}=\eta^{ij}$
to lower or raise the indices $i, j, k\cdots$,
which run over 0,1,2,3.
The transformation (\ref{eqn:3-6})
is reduced to
\beq
{\mathsf X}\to{\mathsf X}'
={\mathsf A}{\mathsf X}{\mathsf A}^\dag,\hspace{1cm}
{\mathsf A}\in Spin(2,{\mathbb H}).
\label{eqn:4-9}
\eeq
Writing
\beq
{x'}^i=\Lambda^i_{\;\;j}x^j,\;\;\;
\Lambda^T\eta\Lambda=\eta,
\label{eqn:4-10}
\eeq
we obtain, as in (\ref{eqn:3-8}),
\beq
\Lambda^i_{\;\;j}\Gamma_i
={\mathsf A}\Gamma_j{\mathsf A}^\dag,
\label{eqn:4-11}
\eeq
where $\Lambda^i_{\;\;j}=\Lambda^i_{\;\;j}({\mathsf A})$.
That is, the Lorentz tranformations
are represented by real quaternions only,
without recourse to
an imaginary time coordinate.
This is in sharp contrast to
the conventional approach
to the application
of quaternions
to relativity,
which relies heavily upon the scheme
of complex quaternions
arising from the fact
that the Minkowski space-time $M_4$
is formally converted into
Euclidean space ${\mathbb R}^4$ in terms of
the imaginary time coordinate
and that
the division algebra, ${\mathbb H}$,
of real quaternions
is regarded as ${\mathbb R}^4$.
We shall see, however, in the next section
that
the imaginary unit commuting with
Hamilton's units
enters into the theory
for an entirely different reason
not connected to the geometrical structure
of Minkowski space-time.
\\[2mm]
\ind
For $x^2\ne 0$, (\ref{eqn:4-7})
leads to
\beq
{\mathsf X}^{-1}
=\frac 1{-x^2}{\tilde {\mathsf X}},\;\;\;
{\tilde {\mathsf X}}\equiv x^i{\tilde\Gamma}_i,
\label{eqn:4-12}
\eeq
where ${\tilde\Gamma}_i$ is given by (\ref{eqn:3-10}).
Therefore, we have
\beq
\Lambda^i_{\;\;j}{\tilde\Gamma}_i
={\mathsf A}^{\dag\,-1}{\tilde\Gamma}_j{\mathsf A}^{-1}.
\label{eqn:4-13}
\eeq
Restricting the indices as
$I=i$ and $J=j$ in (\ref{eqn:3-12}),
we have\footnote{The quaternionic Dirac algebra
formulated by Rotelli \cite{7)}
is obtained by setting ${\mathsf \gamma}_i=
\sigma_3{\tilde\Gamma}_i
=\Gamma_i\sigma_3$, so that
(\ref{eqn:4-11}) and (\ref{eqn:4-13}) 
read $\Lambda^i_{\;\;j}{\mathsf \gamma}^j=
{\mathsf A}^{-1}{\mathsf \gamma}^i{\mathsf A}$,
and the relation (\ref{eqn:4-14}) becomes
${\mathsf \gamma}_i{\mathsf \gamma}_j+
{\mathsf \gamma}_j{\mathsf \gamma}_i=-2\eta_{ij}$.
}
\beq
\Gamma_i{\tilde\Gamma}_j
+\Gamma_j{\tilde\Gamma}_i=-2\eta_{ij}1_2,\;\;\;
{\tilde\Gamma}_i\Gamma_j
+{\tilde\Gamma}_j\Gamma_i=-2\eta_{ij}1_2.
\label{eqn:4-14}
\eeq
Then
$\Lambda^T\eta\Lambda=\eta$
follows, leading to the conclusion
$\Lambda({\mathsf A})\in SO(3, 1)$,
because det$\,\Lambda=1$.
We already know that
$\Lambda^0_{\;\;0}\ge 1$.\footnote{In the present
case, it is easy to prove this directly
from the following equation
(\ref{eqn:4-15}):
$\Lambda^0_{\;\;0}=(1/2)$Tr$\,[{\mathsf A}{\mathsf A}^\dag]
=|Q|^2+|P|^2=
|Q|^2-|P|^2+2|P|^2
=1+2|P|^2\ge 1$.}
The equation
(\ref{eqn:3-13})
is simplified to
\beq
\Lambda^i_{\;\;j}=-\frac 12\mbox{Tr}\,[
{\mathsf A}\Gamma_j{\mathsf A}^\dag
{\tilde\Gamma}^i].
\label{eqn:4-15}
\eeq
The two-valuedness $\Lambda({\mathsf A})=\Lambda(-{\mathsf A})$
and
the homomorphism $\Lambda^i_{\;\;j}({\mathsf A}_1)
      \Lambda^j_{\;\;k}({\mathsf A}_2)=\Lambda^i_{\;\;k}
      ({\mathsf A}_1{\mathsf A}_2)$
are also
obvious from the result in the previous section
and (\ref{eqn:4-4}).
Hence, 
the following isomorphism is proved:
\beq
Spin(2,{\mathbb H})/Z_2{\cong}L_+^\uparrow.
\label{eqn:4-16}
\eeq
\ind
To summarize, we recall that,
according to
the spinor analysis, 
the proper orthochronous Lorentz transformations
are represented by four complex numbers,
$\alpha,\beta,\gamma$ and $\delta,$
with one complex constraint,
$\alpha\delta-\beta\gamma=1$. Thus there are
six free parameters.
We have found another
representation
of the proper orthochronous Lorentz transformations
in terms of two real quaternions, $Q$ and $P$,
with two real constraints,
$|Q|^2-|P|^2=1$ and Re$\,[{\bar Q}P]=0$.
Thus, again, there are six free parameters.
This leads us to believe that
there must be a connection
between the two representations.
Indeed, this connection is found in \S 6.
\\[2mm]
\ind
In order to express ${\mathsf A}\in Spin(2,{\mathbb H})$ in terms of
$\Lambda^i_{\;\;j}$, we multiply (\ref{eqn:4-11})
with ${\tilde\Gamma}^j$ from right and sum over
$j=0,1,2,3,$ obtaining
\beq
\Lambda^i_{\;\;j}\Gamma_i{\tilde\Gamma}^j
={\mathsf A}\Gamma_j{\mathsf A}^\dag{\tilde\Gamma}^j.
\label{eqn:4-17}
\eeq
The following relations can be shown:
\beq
\Gamma_j{\mathsf A}^\dag{\tilde\Gamma}^j=-4{\mbox{Re}}\,
{\mathsf A}^T,\;\;\;
({\mbox{Re}}\,
{\mathsf A}^T)^{-1}=\frac 1{Q_0^2+P_0^2}{\mbox {Re}}
\,{\mathsf A},\;\;\;
Q_0^2+P_0^2=\frac 14\mbox{tr}\,\Lambda,\;\;\;
\mbox{tr}\,\Lambda=\Lambda^i_{\;\;i}.
\label{eqn:4-18}
\eeq
Substituing thes results into (\ref{eqn:4-17})
and using the relation 
Re$\,[\Gamma_i{\tilde\Gamma}^j]=-\eta_i^{\;\;j}1_2$,
we finally obtain
\beq
{\mathsf A}=-\frac 1{\mbox{tr}\,\Lambda}
(\Lambda^i_{\;\;j}\Gamma_i{\tilde\Gamma}^j)
({\mbox {Re}}\,{\mathsf A}).
\label{eqn:4-19}
\eeq
This equation gives Im$\,{\mathsf A}$
for given $\Lambda^i_{\;\;j}$ and Re$\,{\mathsf A}$.
At first sight, it seems to be useless, because
it does not reveal how the matrix $\,{\mathsf A}$ itself
is determined once $\Lambda^i_{\;\;j}$ is given, as
is the case in the spinor analysis.
Nevertheless, we can extract a meaningful result from it
if we have
\beq
{\mathsf A}=1_2+{\mathsf \Sigma},\;\;\;
{\overline {\mathsf \Sigma}}=-{\mathsf \Sigma}.
\label{eqn:4-20}
\eeq
This is the case for infinitesimal transformations,
$\Lambda^i_{\;\;j}=\eta^i_{\;\;j}+\omega^i_{\;\;j}$
with $\omega_{ij}=-\omega_{ji}$,
where ${\bar{\mathsf A}}{\mathsf A}=(
1_2+{\overline {\mathsf \Sigma}})(1_2+{\mathsf \Sigma})
=1_2+{\overline {\mathsf \Sigma}}+{\mathsf \Sigma}=1_2$
for infinitesimal ${\mathsf \Sigma}$.
We then find from (\ref{eqn:4-19}) that
\beq
{\mathsf \Sigma}=-\frac 14\omega_{ij}\Sigma^{ij},\;\;\;
\Sigma^{ij}=\frac 12(\Gamma^i{\tilde\Gamma}^j-
\Gamma^j{\tilde\Gamma}^i).
\label{eqn:4-21}
\eeq
The generators $\Sigma^{ij}$ satisfy the commutation relations
\beq
[\Sigma^{ij},\Sigma^{kl}]
=2[\eta^{ik}\Sigma^{jl}
-\eta^{il}\Sigma^{jk}
-\eta^{jk}\Sigma^{il}
+\eta^{jl}\Sigma^{ik}].
\label{eqn:4-22}
\eeq
These are the same commutation relations
obeyed by the Lorentz generators
$a^{ij}$ which are defined by
$\omega^i_{\;\;j}x^j
=-(1/4)\omega_{kl}(a^{kl})^i_{\;\;j}x^j$.
Hence, $Spin(2,{\mathbb H})$
is locally isomorphic to $L_+^\uparrow$.
\\[2mm]
\ind
The generators in the 3-dimensional notation
can be obtained from the formulae
given in Appendix A. Those for spatial rotations
are given by
\beq
\Lambda^a=
\left(
\barr{cc}
e_a&0\\
0&e_a\\
\earr
\right),
\;\;\;
a=1,2,3,
\label{eqn:4-23}
\eeq
which are anti-Hermitian,
while the boost generators are merely Hermitian
$\Gamma^a,\;a=1,2,3$.
Their commutation relations are 
\beq
[\Lambda^a, \Lambda^b]=2\epsilon_{abc}\Lambda^c,\;\;\;
[\Lambda^a, \Gamma^b]=2\epsilon_{abc}\Gamma^c,\;\;\;
[\Gamma^a, \Gamma^b]=-2\epsilon_{abc}\Lambda^c.
\label{eqn:4-24}
\eeq
Note here that $\Gamma^a=\Sigma^{a0}$ and
$\Sigma^{ab}=\epsilon_{abc}\Lambda^c$.
\section{Dirac theory based on $Spin(2,{\mathbb H})$} 
It is well known that the spinor analysis, which originated from
the Dirac theory, is based upon the isomorphism
\beq
SL(2,{\mathbb C})/Z_2{\cong}L_+^\uparrow.
\label{eqn:5-1}
\eeq
The fundamental spinors $\xi^\alpha\;(\alpha=1,2)$
and ${\bar\eta}_{{\dot\alpha}}
\;({\dot \alpha}={\dot 1},{\dot 2})$,
which transform as
\beqa
{\xi'}^\alpha&=&A^\alpha_{\;\;\beta}\xi^\beta,\;\;\;
(A)_{\alpha\beta}\equiv A^\alpha_{\;\;\beta},\;\;\;
A\in SL(2,{\mathbb C}),\nn\\[2mm]
{{\bar\eta}'}_{{\dot \alpha}}&=&
A_{\dot \alpha}^{\;\;{\dot \beta}}
{\bar\eta}_{{\dot \beta}},\;\;\;
(A^{\dag\,-1})_{\alpha\beta}
\equiv A_{\dot \alpha}^{\;\;{\dot \beta}},
\label{eqn:5-2}
\eeqa
constitute the Dirac spinor (in the Weyl representation),
\beq
\psi
=
\left(
\barr{l}
\xi^\alpha\\
{\bar\eta}_{{\dot \alpha}}\\
\earr
\right),
\label{eqn:5-3}
\eeq
which undergoes the Lorentz transformation
\beq
\psi'(x')
=
\left(
\barr{cc}
A&0\\
0&A^{\dag\,-1}\\
\earr
\right)
\psi(x),\;\;\;
x'=\Lambda x,
\label{eqn:5-4}
\eeq
with
$\Lambda^i_{\;\;j}\sigma_i=A\sigma_jA^\dag$.
Thanks to Pauli's lemma, one can introduce
the Dirac spinor $\psi$ in any representation
of the Dirac matrix, $\gamma^i\;(i=0,1,2,3)$,
satisfying
\beq
\gamma^i\gamma^j+\gamma^j\gamma^i=2\eta^{ij},
\label{eqn:5-5}
\eeq
where $\psi$ undergoes the Lorentz transformation 
\beq
\psi'(x')
=
S(\Lambda)\psi(x),\;\;\;
S^{-1}(\Lambda)\gamma^iS(\Lambda)
=\Lambda^i_{\;\;j}\gamma^j.
\label{eqn:5-6}
\eeq
\ind
On the basis of the isomorphism
(\ref{eqn:4-16}),
we find another formulation
of the Dirac theory regarding a quaternionic 2-component object
${\mathsf \Psi}$
as a Dirac spinor
subject to the linear Lorentz transformation law
\beqa
{\mathsf \Psi}(x)&=&
\left(
\barr{l}
{\bf \zeta}^1(x)\\
{\bf \zeta}^2(x)\\
\earr
\right),\;\;\;{\bf \zeta}^1(x), {\bf \zeta}^2(x)\in{\mathbb H},\nn\\[2mm]
{\mathsf \Psi}'(x')&=&{\mathsf A}{\mathsf \Psi}(x),
\;\;\;{\mathsf A}\in Spin(2,{\mathbb H}).
\label{eqn:5-7}
\eeqa
It is assumed that ${\mathsf \Psi}$ is a Grassmannian quaternion:
\beq
{\mathsf \Psi}=e_i{\mathsf \Psi}_i(i=0,1,2,3),
\;\;\;
{\mathsf \Psi}_i{\mathsf \Psi}_j=-
{\mathsf \Psi}_j{\mathsf \Psi}_i.
\label{eqn:5-8}
\eeq
\ind
If ${\mathsf A}$ is a general element ${\mathbf A}$
of
$SL(2,{\mathbb H})$,
${\mathsf \Psi}$ is called a $SL(2,{\mathbb H})$ spinor,
which is a Weyl spinor
in $D=6$ space-time.
Such a $SL(2,{\mathbb H})$ spinor was considered by G\"ursey \cite{10)}
but its kinematics, together with another Weyl spinor
in $D=6$,
was thoroughly investigated by Kugo and Townsend, \cite{3)}
who pointed out the relation
between the division algebras, 
${\mathbb R}, {\mathbb C}, {\mathbb H}$ (and possibly ${\mathbb O}$), 
and the supersymmetry in
$D=3,4,6$ (and possibly in $D=10$), respectively,
and, in particular,
presented a detailed description 
of $D=6$ supersymmetry in terms of the $SL(2,{\mathbb H})$ spinor.
\\[2mm]
\ind
Our ${\mathsf \Psi}$ is not a $SL(2,{\mathbb H})$ spinor
but, rather, the Dirac spinor in quaternion language.
As such, it is not expected that
anything new appears.
Nonetheless, it would be inetersting to see
what kind of formalism is needed
by regarding ${\mathsf \Psi}$ as a Dirac spinor
in $D=4$ but not as a Weyl spinor in $D=6$, as in
Ref. 3).
It turns out that
Dirac field operators are automatically 
`anti-symmetrized'
due to the fact
that
the charge conjugation
transformation for ${\mathsf \Psi}$
becomes linear, in contrast to
the non-linear charge conjugation transformation
in the conventional treatment.
An investigaiton along this line by the
present author is given in Ref.~4).
The present section is largely based on that work,
but here we add detailed manipulations
with emphasis on the important role played
by the Pauli-G\"ursey group. \cite{8)}
\\[2mm]
\ind
Each component of ${\mathsf \Psi}$ must satisfy the 
Klein-Gordon equation
\beq
(\Box-m^2){\mathsf \Psi}=0,
\label{eqn:5-9}
\eeq
with $\Box=\eta^{ij}\partial_i\partial_j$.
This equation is obtained from the
following quaternionic Dirac 
equation: \cite{4)}$^{,}$\footnote{The same
quaternionic Dirac equation was obtained independently
by Rotelli \cite{7)}. In terms of Rotelli's quaternionic
Dirac matrix ${\mathsf \gamma}^i$
defined in the footnote on p.15, it reads
$\gamma^i\partial_i{\mathsf \Psi}-m{\mathsf \Psi}e_3=0$.}\
\beq
{\tilde\Gamma}^i\partial_i{\mathsf \Psi}
-m\sigma_3{\mathsf \Psi}e_3=0.
\label{eqn:5-10}
\eeq
To prove this, we
operate on this equation
with $\Gamma^j\partial_j$ from left
to obtain
\bea*
\Gamma^j\partial_j{\tilde\Gamma}^i\partial_i{\mathsf \Psi}
-m\Gamma^j\partial_j\sigma_3{\mathsf \Psi}e_3
=\frac 12(\Gamma^j{\tilde\Gamma}^i+\Gamma^i{\tilde\Gamma}^j)
\partial_i\partial_j{\mathsf \Psi}
-m\Gamma^j\partial_j\sigma_3{\mathsf \Psi}e_3=0.
\eea*
Using (\ref{eqn:4-14})
and noting the relation 
$\Gamma^i\sigma_3=\sigma_3{\tilde\Gamma}^i\;(i=0,1,2,3)$, we
arrive at
\bea*
-\Box{\mathsf \Psi}
-m\sigma_3{\tilde\Gamma}^j\partial_j{\mathsf \Psi}e_3
=-\Box{\mathsf \Psi}-(m\sigma_3)(m\sigma_3{\mathsf \Psi}
e_3)e_3
=(-\Box+m^2){\mathsf \Psi}=0.
\eea*
In this derivation,
the presense of the imaginary unit $e_3$
on the right in the mass term of (\ref{eqn:5-10})
is particulary important.\footnote{The imaginary unit itself
may be chosen as $e_1, e_2$ or $e_3$ but we keep our 
definition for convenience.}\
The matrix $\sigma_3$ on the
left in the mass term of (\ref{eqn:5-10})
converts the matrix $\Gamma^i$ to
${\tilde\Gamma}^i$ via the relation alluded to above.
The matrix $\sigma_1$ may also be employed for this
purpose.
\\[2mm]
\ind
The Dirac equation (\ref{eqn:5-10})
is Lorentz invariant
with the transformation law
${\mathsf \Psi}(x)\to
{\mathsf \Psi}'(x')
={\mathsf A}{\mathsf \Psi}(x)$
for ${\mathsf A}\in Spin(2,{\mathbb H})$,
because of (\ref{eqn:4-13}), 
and possesses the Abelian smmetry,
${\mathsf \Psi}\to{\mathsf \Psi}e^{\alpha e_3}$ for 
global $\alpha$.
It also possesses $P, C$ and $T$ invariances; the discrete
transformations (in the one-particle theory)
are defined by
${\mathsf \Psi}(x)\to{\mathsf \Psi}'(x_P)=
\pm\sigma_3{\mathsf \Psi}(x),
{\mathsf \Psi}(x)\to {\mathsf \Psi}^c(x)=
-\omega{\mathsf \Psi}(x)e_1,
{\mathsf \Psi}(x)\to
{\mathsf \Psi}'(x_T)=
-\sigma_3{\mathsf \Psi}(x)e_2$,
respectively,
under $P, C$ and $T$, where
$x_P=(x^0, -x^1,$\\
$-x^2, -x^3)$
and
$x_T=(-x^0, x^1, x^2, x^3)$.
It should be noted that
the charge conjugation transformation
is linear, while the usual
4-component formalism assumes the non-linear
$C$-transformation property $\psi\to
C{\tilde\psi}^T, {\tilde\psi}=\psi^\dag\gamma^0$,
with $C$ the charge conjugation matrix. In the massless
case ($m=0$) it also possesses the chiral symmetry,
${\mathsf \Psi}\to e^{\alpha\omega}{\mathsf \Psi}$
for global $\alpha$,
and an extra non-Abelian symmetry,
${\mathsf \Psi}\to {\mathsf \Psi}q$,
where $q$ is a unit quaternion.
We require below that this non-Abelian symmetry,
which
gives rise to the Pauli-G\"ursey symmetry,
remain intact in the Lagrangian formalism.
It turns out that
this requirement has a non-trivial consequence.
\\[2mm]
\ind
The momentum-space expansion of ${\mathsf \Psi}(x)$
is given by \footnote{The sum over the momentum
is given by
$\sum_{{\scriptsize{\bp}}}\equiv
\int\frac {d^3{\scriptsize{\bp}}}
{(2\pi)^3E({\scriptsize{\bp}})},\;\;$with
$E(\bp)=\sqrt{\bp^2+m^2}$.}\
\beq
{\mathsf \Psi}(x)
=\sum_{{\scriptsize{\bp}},r=\pm}
[{\mathsf U}^{(r)}(p)A(\bp, r)e^{-(px)e_3}
+{\mathsf V}^{(r)}(p){\bar B}(\bp, r)e^{(px)e_3}],
\label{eqn:5-11}
\eeq
where $A$ and $B$ are complex Grassmannian numbers
over ${\mathbb C}(1, e_3)$ given by
\beq
A=\alpha_r-\alpha_ie_3,\;\;\;
B=\beta_r-\beta_ie_3,
\label{eqn:5-12}
\eeq
with real Grassmannian numbers, $\alpha_r, \alpha_i,
\beta_r$ and $\beta_i$.
The imaginary unit $e_3$
should be placed on the right
of the spinors ${\mathsf U}^{(r)}(p)$ 
and ${\mathsf V}^{(r)}(p)$, which satisfy the
$e_3$-free Dirac equation
\beq
({\tilde\Gamma}^ip_i+m\sigma_3){\mathsf U}^{(r)}(p)=0,
\;\;\;
({\tilde\Gamma}^ip_i-m\sigma_3){\mathsf V}^{(r)}(p)=0.
\label{eqn:5-13}
\eeq
Since det$\,({\tilde\Gamma}^ip_i+m\sigma_3)
=$det$\,({\tilde\Gamma}^ip_i-m\sigma_3)=(p^2+m^2)^2$,
we have $p^2+m^2=0$.
We set $p^0=\sqrt{|\bp|^2+m^2}=E(\bp)>0$,
as usual.
\\[2mm]
\ind
Defining the `helicity' operator
\beq
s(\bp)=\frac{{\bf\Lambda}\cdot\bp}
{\sqrt{|\bp|^2}}
\label{eqn:5-14}
\eeq
in terms of the spatial rotation generators (\ref{eqn:4-23}),
we have only two independent solutions, since $s^2(\bp)=-1_2$:
\beq
s(\bp){\mathsf U}^{(\pm)}(p)=\pm {\mathsf U}^{(\pm)}(p)e_3,\;\;\;
s(\bp){\mathsf V}^{(\pm)}(p)=\mp {\mathsf V}^{(\pm)}(p)e_3.
\label{eqn:5-15}
\eeq
The orthonormality conditions for the spinors take the forms
\beq
\mbox{Re}\,[{{\mathsf U}^{(r)}}^\dag(p){\mathsf U}^{(s)}(p)]
=2E(\bp)\delta_{rs},\;\;\;
\mbox{Re}\,[{{\mathsf V}^{(r)}}^\dag(p){\mathsf V}^{(s)}(p)]
=2E(\bp)\delta_{rs},
\label{eqn:5-16}
\eeq
${\mathsf U}^{(r)}(p)$ and ${\mathsf V}^{(s)}(p)$
being orthogonal to each other.
Similarly, the completeness conditions
are
\beq
\sum_r{\mathsf U}^{(r)}(p){{\mathsf U}^{(r)}}^\dag(p)
=2(-\Gamma^ip_i+m\sigma_3),\;\;\;
\sum_r{\mathsf V}^{(r)}(p){{\mathsf V}^{(r)}}^\dag(p)
=2(-\Gamma^ip_i-m\sigma_3).
\label{eqn:5-17}
\eeq
\ind
In order to carry out a second quantization of the theory,
we postulate the following(omitting the indices $\bp, r$, etc.)
\beq
\{A,{\bar A}\}=1,\;\;\;\{B,{\bar B}\}=1,
\label{eqn:5-18}
\eeq
with all other anticommutators vanishing.
We see below in this section
that
the requirement
that
the Pauli-G\"ursey symmetry group
$SU(2)$ in the massless Dirac Lagrangian
remain intact also in the Hermitian
Dirac Lagrangian constructed from ${\mathsf \Psi}$
leads to the inevitable introduction into the theory
of an additional imaginary unit
that commutes with Hamilton's units.
Taking this additinonal imaginary unit
to be the ordinary imaginary unit $i$,
we are free to
make use of $i$
also in the second quantization
of the theory.
Our approach is, therefore, essentially different
from that employed in the quaternionic quantum mechanics
and quaternionic quantum field theory,
where such an introduction
of $i$ into the quaternionic theory
is strictly avoided. \cite{12)}
\\[2mm]
\ind
Because we are free to make use of $i$, we introduce
the projection operators
\beq
E_{\pm}=\frac 12(e_0\pm ie_3),
\label{eqn:5-19}
\eeq
where $i$ is assumed to commute with Hamilton's units 
$e_a\;(a=1,2,3)$.
The projection operators in (\ref{eqn:5-19})
satisfy
\beq
E^2_{\pm}=
E_{\pm},\;\;\;E_+E_-=E_-E_+=0,\;\;\;
e_3E_{\pm}=\mp iE_\pm.
\label{eqn:5-20}
\eeq
Consequently, we have 
\beqa
&&(A,{\bar A},B,{\bar B})E_+=
(a,a^\dag, b, b^\dag)E_+,
\;\;\;
(A,{\bar A},B,{\bar B})E_-=
(a^\dag,a, b^\dag, b)E_-,\nn\\[2mm]
&&a\equiv\alpha_r+i\alpha_i,
\;\;\;
a^\dag\equiv\alpha_r-i\alpha_i,\;\;\;
b\equiv\beta_r+i\beta_i,
\;\;\;
b^\dag\equiv\beta_r-i\beta_i,
\label{eqn:5-21}
\eeqa
which reproduce the usual anti-commutation
relations among the operators (\ref{eqn:5-21})
from the ansatz (\ref{eqn:5-18}).
\\[2mm]
\ind
By definiton, $E_++E_-=e_0$ is the unit element of ${\mathbb H}$,
and thus we have the relation
\beq
{\mathsf \Psi}={\mathsf \Psi}_++{\mathsf \Psi}_-,\;\;\;
{\mathsf \Psi}_\pm\equiv{\mathsf \Psi}E_\pm.
\label{eqn:5-22}
\eeq
It follows from the last equation of (\ref{eqn:5-20})
that each ${\mathsf \Psi}_\pm$ has an ordinary separation
of positive- and negative-frequency parts,
$e^{ipx}$ and $e^{-ipx}$, respectively,
with particle (anti-particle) annihilation operators $a$ ($b$) 
and anti-particle (particle)
creation operators $b^\dag$ ($a^\dag$), respectively,
for ${\mathsf \Psi}_+$ (${\mathsf \Psi}_-$).
\\[2mm]
\ind
Since ${\mathsf \Psi}_-$ is the complex conjugate
(the Hermitian conjugate in operator language)
of ${\mathsf \Psi}_+$,
if we assume that
${\mathsf \Psi}_+$ is an (internal-symmetry)
multiplet 
belonging
to ${\underline{n}}$ of $SU(n)$,
${\mathsf \Psi}_-$ belongs to
${\underline {n}}^*$ of $SU(n)$,
which is not equivalent to
${\underline{n}}$ unless $n=2$.
The left-translation
(including the case $n=2$)
and the right-translation
common to both ${\mathsf \Psi}_\pm$
can be treated without considering
the above
decomposition.
In particular,
the Lorentz transformation
is a left-translation
common to both ${\mathsf \Psi}_\pm$.
\\[2mm]
\ind
We now go on to find the Dirac Lagrangian
which leads to (\ref{eqn:5-10}) employing
the action principle.
Since Lagrangian must be Hermitian,
we choose it as\footnote{The Hermiticity
requirement means that
the Lagrangian is the real part of a quaternion.
If this assumption is not made,
there is no hope to maintain the symmetry ${\mathsf \Psi}\to
{\mathsf \Psi}q$ for a unit quaternion $q$
already present in the equation level (\ref{eqn:5-10})
with $m=0$
in the Lagrangian formalism.
For this reason the Hermitian Dirac Lagrangian
proposed in Ref. 13), which is purely imaginary,
is not accepted.}\
\bea*
\cL'=\mbox{Re}\,[e_3{\mathsf \Psi}^\dag
{\tilde\Gamma}^i\partial_i
{\mathsf \Psi}-me_3{\mathsf \Psi}^\dag\sigma_3{\mathsf \Psi}e_3],
\eea*
where Hermitian conjugation reverses the order of
the operators, together 
with
$e_a\to{\bar e_a}=-e_a\;(a=1,2,3)$.
The above Lagrangian is Hermitian 
up 
to a total divergence,\footnote{The surface term
in the action vanishes if we assume,
for instance,
periodic boundary conditions.
It should be noted here that the above mass term identically vanishes
due to the Grassmannian nature of the spinor ${\mathsf \Psi}$.
This difficulty is resolved through the introduction of $i$.} as we have
\bea*
\cL{'}^\dag=\mbox{Re}\,[\partial_i{\mathsf \Psi}^\dag
{\tilde\Gamma}^i
{\mathsf \Psi}{\bar e}_3-m
{\bar e}_3{\mathsf \Psi}^\dag\sigma_3{\mathsf \Psi}
{\bar e}_3]
=\mbox{Re}\,[e_3{\mathsf \Psi}^\dag
{\tilde\Gamma}^i\partial_i
{\mathsf \Psi}-me_3{\mathsf \Psi}^\dag\sigma_3{\mathsf \Psi}e_3]
=\cL'.
\eea*
In addition to the chiral
symmetry,
the massless Dirac Lagrangian
has a global non-Abelian symmetry,
the Pauli-G\"ursey $SU(2)$ symmetry, \cite{8)}
which is not an internal symmetry
related to a multiplet structure
of the spinor ${\mathsf \Psi}$.
Since $SU(2)\cong
SL(1,{\mathbb H})=\{q\in{\mathbb H};\;|q|=1\}$, the kinetic energy part of
the Dirac Lagrangian that we seek
must be invariant under
${\mathsf \Psi}\to {\mathsf \Psi}q$
for $q\in SL(1,{\mathbb H})$, because the quaternionic
Dirac equation (\ref{eqn:5-10})
possesses the same symmetry.
However, this is impossible for the above
choice of the Lagrangian
because quaternions are not commutative.
\\[2mm]
\ind
If we can assume that the left-most $e_3$
in the kinetic energy term of the
above Lagrangian commutes with
quaternions, we can construct a
Hermitian Lagrangian
with symmetry under the right-translation.
We are thus forced to replace the left-most $e_3$
in $\cL'$ with an additional
imaginary unit which commutes with
quaternions and changes sign under 
Hermitian conjugation.\footnote{The left-most $e_3$
in the mass term of the above Lagrangian 
should also be replaced by $i$ to reproduce
the 
Dirac equation (\ref{eqn:5-10}). Then the mass term
is proportional to Re$\,[{\mathsf \Psi}^\dag
\sigma_3{\mathsf \Psi}e_3]$, which is now nonvanishing.
The overbar on the second spinor factor 
in the first term of (26) in Ref. 4)
should be deleted.}
We take it as the ordinary imaginary unit
$i$ used already in the second quantization
to finally obtain 
\beq
\cL=i\mbox{Re}\,[{\mathsf \Psi}^\dag
{\tilde\Gamma}^i\partial_i
{\mathsf \Psi}-m{\mathsf \Psi}^\dag\sigma_3{\mathsf \Psi}e_3].
\label{eqn:5-23}
\eeq
We stress that
the relativistic Dirac equation (\ref{eqn:5-10})
never requires the introduction
of $i$ into the theory.
Hence we conclude that
$i$ comes from the requirement
that the Dirac theory
based on the spinor ${\mathsf \Psi}$
allow a consistent Lagrangian formalism.\\[2mm]
\ind
Noting that,
for any Grassmannian quaternions $p=e_ip_i,q=e_iq_i$
with $\{p_i, p_j\}=\{q_i, q_j\}=\{p_i, q_j\}=0$, we have
\beqa
{\overline{qp}}
&=&-{\bar p}{\bar q},\nn\\[2mm]
\mbox{Re}\,
[qp]&=&-\mbox{Re}\,[pq],
\label{eqn:5-24}
\eeqa
the variation of (\ref{eqn:5-23})
is given by (up to a total divergence)
\bea*
\delta\cL&=&i\mbox{Re}\,[{\delta{\mathsf \Psi}}^\dag
{\tilde\Gamma}^i\partial_i{\mathsf \Psi}
+{\mathsf \Psi}^\dag
{\tilde\Gamma}^i\partial_i\delta{\mathsf \Psi}
-m\delta{\mathsf \Psi}^\dag\sigma_3{\mathsf \Psi}e_3
-m{\mathsf \Psi}^\dag\sigma_3\delta{\mathsf \Psi}e_3]\nn\\[2mm]
&=&i\delta{\mathsf \Psi}_i
\mbox{Re}\,[{\bar e}_i{\tilde\Gamma}^j
\partial_j{\mathsf \Psi}
+\partial_j{\mathsf \Psi}^\dag{\tilde\Gamma}^je_i
-m{\bar e}_i\sigma_3{\mathsf \Psi}e_3
+m{\mathsf \Psi}^\dag\sigma_3e_i
e_3]
\nn\\[2mm]
&=&
i\delta{\mathsf \Psi}_i
\mbox{Re}\,[{\bar e}_i{\tilde\Gamma}^j\partial_j{\mathsf \Psi}
+{\bar e}_i{\tilde\Gamma}^j
\partial_j{\mathsf \Psi}
-m{\bar e}_i\sigma_3{\mathsf \Psi}e_3
-me_3{\bar e}_i\sigma_3{\mathsf \Psi}]\nn\\[2mm]
&=&
2i\mbox{Re}\,[\delta{\mathsf \Psi}^\dag
({\tilde\Gamma}^i\partial_i{\mathsf \Psi}
-m\sigma_3{\mathsf \Psi}e_3)].
\eea*
We have thus shown\footnote{The above manipulation
indicates that the variation
$\delta{\mathsf \Psi}$ is not indpendent
of $\delta{\overline{\mathsf \Psi}}$.
This is because, if we define the quaternionic deivative
$\partial/\partial q$
such that
$\partial q/\partial q=1$, 
we necessarily have
$\partial {\bar q}/\partial q=-2$.
This contrasts with
the complex case, in which 
$\partial z/\partial z=1$
leads to
$\partial z^*/\partial z=0$.
(See, however, Lemma 3 on p. 375 of Ref.~12).)}
that the
action principle applied to
the Dirac Lagrangian (\ref{eqn:5-23}),
$\delta\!\int d^4x\cL=0$ for 
arbitrary $\delta{\mathsf \Psi}$
with vanishing surface term,
yields the Dirac equation (\ref{eqn:5-10}). 
\\[2mm]
\ind
Thanks to the presence
of $i$ in the theory,
we can make contact with the conventional formalism
in terms of the 4-component Dirac spinor
by using
the Pauli representation
$\rho$ (2$\cdot$15) of ${\mathbb H}$:
\beq
\rho({\mathsf \Psi})=
(\psi, \Omega\psi^*),\;\;\;
\rho({\mathsf \Psi}^\dag)=
\left(
\barr{c}
\psi\dag\\
-\psi^T\Omega
\earr
\right).
\label{eqn:5-25}
\eeq
Here, $\rho({\mathsf \Psi})$ and
$ \rho({\mathsf \Psi}^\dag)$ are, respectively,
$4\times 2$ and 2$\times$ 4 matrices, with
\beq
\psi=
\left(
\barr{l}
\zeta^1_0-i\zeta^1_3\\
-i\zeta^1_1+\zeta^1_2\\
\zeta^2_0-i\zeta^2_3\\
-i\zeta^2_1+\zeta^2_2\\
\earr
\right),\;\;\;
\psi^*=
\left(
\barr{l}
\zeta^1_0+i\zeta^1_3\\
+i\zeta^1_1+\zeta^1_2\\
\zeta^2_0+i\zeta^2_3\\
+i\zeta^2_1+\zeta^2_2\\
\earr
\right),\;\;\;
\Omega=\left(
\barr{cc}
\omega&0_2\\
0_2&\omega\\
\earr
\right),
\label{eqn:5-26}
\eeq
where $0_2$ is the 2-dimensional null matrix
and $\zeta^a=e_0\zeta^a_0+
e_1\zeta^a_1+e_2\zeta^a_2+e_3\zeta^a_3\;
(a=1,2)$. 
The momentum-space expansion
(\ref{eqn:5-11}) corresponds to the familiar
expansion
\beq
\psi(x)
=\sum_{{\scriptsize{\bp}},r=\pm}
[u^{(r)}(p)a(\bp, r)e^{i(px)}
+v^{(r)}(p)b^\dag(\bp, r)e^{-i(px)}],
\label{eqn:5-27}
\eeq
and a similar one for $\Omega\psi^*$,
where the operators $a$ and $b^\dag$ are defined
by (\ref{eqn:5-21}).\\[2mm]
\ind
We next set
\beq
\rho({\tilde\Gamma}^i)
=-\gamma^0\gamma^i,\;\;\;(i=0,1,2,3)
\label{eqn:5-28}
\eeq
with
\beq
\gamma^0=
\left(
\barr{cc}
-i1_2&0_2\\
0_2&i1_2\\
\earr
\right),\;\;\;
\gamma^a=
\left(
\barr{cc}
0_2&-\sigma^a\\
-\sigma^a&0_2\\
\earr
\right)
,\;\;\;\sigma^a=\sigma_a,
\;\;\;(a=1,2,3)
\label{eqn:5-29}
\eeq
and thereby obtain
\beq
\rho(\Gamma^i)
=-\gamma^i\gamma^0.\;\;\;(i=0,1,2,3)
\label{eqn:5-30}
\eeq
Hence (\ref{eqn:4-14}) corresponds to the
Dirac algebra
\beq
-\rho(\Gamma^i{\tilde\Gamma}^j
+\Gamma^j{\tilde\Gamma}^i)
=\gamma^i\gamma^j+\gamma^j\gamma^i
=2\eta_{ij}\rho(1_2)
=2\eta_{ij}1_4.\;\;\;(i,j=0,1,2,3)
\label{eqn:5-31}
\eeq
We call (\ref{eqn:5-29})
the Pauli representation
of the gamma matrices. 
The Dirac equation
(\ref{eqn:5-10}) reads
\beqa
\rho({\tilde\Gamma}^i\partial_i{\mathsf \Psi}
-m\sigma_3{\mathsf \Psi}e_3)
&=&-\gamma^0\big(\gamma^i\partial_i(\psi,\Omega\psi^*)
+m(\psi,-\Omega\psi^*)\big)=0,\nn\\[2mm]
(\gamma^i\partial_i+m)\psi&=&0,\;\;\;
{\tilde\psi}(\gamma^i\overleftarrow{\partial_i}
-m)=0,\;\;\;{\tilde\psi}\equiv\psi^\dag\gamma^0.
\label{eqn:5-32}
\eeqa
Also, using (\ref{eqn:2-17}) the Dirac Lagrangian (\ref{eqn:5-23})
is found to be
\beqa
\cL
&=&\frac 12
\mbox{tr}\,
\Biggl[
\left(
\barr{c}
\psi^\dag\\
-\psi^T\Omega\\
\earr
\right)
(-i)\gamma^0\gamma^i\partial_i
(\psi, \Omega\psi^*)\nn\\[2mm]
&&\mbox{\hspace{0.8cm}}-im\left(
\barr{l}
\psi^\dag\\
-\psi^T\Omega\\
\earr
\right)
\left(
\barr{cc}
1_2&0_2\\
0_2&-1_2\\
\earr
\right)(-i\psi, i\Omega\psi^*)
\Biggr]\nn\\[2mm]
&=&-\frac i2[
{\tilde\psi}\gamma^i\partial_i\psi
+\psi^T\gamma^{i\,T}\partial_i{\tilde\psi}^T]
-\frac {im}2({\tilde\psi}\psi-\psi^T{\tilde\psi}^T)\nn\\[2mm]
&=&-\frac i2[
{\tilde\psi},(\gamma^i\partial_i+m)\psi]
\;\;\;
(\mbox{up to a total divergence}),
\label{eqn:5-33}
\eeqa
where we have used the relation
$\Omega\gamma^0\gamma^i\Omega\gamma^0=
\gamma^{i\,T}$.
The reason why we automatically
obtained the `anti-symmetrized' form of the
Dirac Lagrangian (\ref{eqn:5-33})
comes from the fact
that
the charge-conjugation
transformation
for the original quaternionic
Dirac spinor ${\mathsf \Psi}$
is defined to be linear, i.e.
${\mathsf \Psi}\to{\mathsf \Psi}^c
=-\omega{\mathsf \Psi}e_1$.\footnote{The Majorana
condition
${\mathsf \Psi}^c
=-\omega{\mathsf \Psi}e_1={\mathsf \Psi}$
implies that
$\zeta^2=-\zeta^1e_1$.
}
To be more precise, we have the representation
\beqa
\rho({\mathsf \Psi}^c)
=(\psi^c, \Omega\psi^{c\,*}),\;\;\;
\psi^c\equiv C{\tilde\psi}^T,
\label{eqn:5-34}
\eeqa
where the charge conjugation matrix $C=\left(
                                     \barr{cc}
                                     0_2&i\sigma_2\\
                                     i\sigma_2&0_2\\
                                     \earr
                                     \right)$
satisfies
$C^{-1}\gamma^iC=-\gamma^{i\,T}$.
Recalling $\rho({\mathsf \Psi})
=(\psi, \Omega\psi^*)$, we recover the usual
charge-conjugation transformation
$\psi\to\psi^c=C{\tilde\psi}^T$,
which is nonlinear.
The parity transformation
is given by
${\mathsf \Psi}(x)\to{\mathsf \Psi}'(x_P)
=\pm\sigma_3{\mathsf \Psi}(x)$,
because
$\rho\big({\mathsf \Psi}'(x_P)\big)
=(\psi'(x_P), \Omega\psi{'}^*(x_P))
=\pm\rho\big(\sigma_3{\mathsf \Psi}(x)\big)
=\pm i(\gamma^0\psi(x),
\gamma^0\Omega\psi^*(x))$.\footnote{We
list here bilinear covariants
and their Pauli representation, in which 
the conventional 4-component spinor
notation is used and which can be converted into
any representation:
\ben
\item[(S)]
Re$\,[{\mathsf \Psi}^\dag\sigma_3{\mathsf \Psi}e_3]$
=(1/2)tr$\,\rho({\mathsf \Psi}^\dag
\sigma_3{\mathsf \Psi}e_3)=
(1/2)[{\tilde\psi},\psi]$;
\item[(P)]
Re$\,[{\mathsf \Psi}^\dag\sigma_1{\mathsf \Psi}e_3]$
=(1/2)tr$\,\rho({\mathsf \Psi}^\dag
\sigma_1{\mathsf \Psi}e_3)=
-(i/2)[{\tilde\psi},\gamma_5\psi]$;
\item[(V)]
Re$\,[{\mathsf \Psi}^\dag
{\tilde{\Gamma}}^i{\mathsf \Psi}e_3]$
=(1/2)tr$\,\rho({\mathsf \Psi}^\dag
{\tilde{\Gamma}}^i{\mathsf \Psi}e_3)=
-(i/2)[{\tilde\psi},\gamma^i\psi]$;
\item[(A)]
Re$\,[{\mathsf \Psi}^\dag
{\tilde{\Gamma}}^i\omega{\mathsf \Psi}]$
=(1/2)tr$\,\rho({\mathsf \Psi}^\dag
{\tilde{\Gamma}}^i{\mathsf \Psi}e_3)=
(i/2)[{\tilde\psi},\gamma_5\gamma^i\psi]$;
\item[(T)]
Re$\,[{\mathsf \Psi}^\dag
{\tilde{\Sigma}}^{ij}\sigma_3{\mathsf \Psi}]$
=(1/2)tr$\,\rho({\mathsf \Psi}^\dag
{\tilde{\Sigma}}^{ij}\sigma_3{\mathsf \Psi})=
(1/2)[{\tilde\psi},\sigma^{ij}\psi]$,
\een
with
${\tilde{\Sigma}}^{ij}
=(1/2)({\tilde{\Gamma}}^i\Gamma^j
-{\tilde{\Gamma}}^j\Gamma^i)$
and
${\sigma}^{ij}
=(1/2i)(\gamma^i\gamma^j
-\gamma^j\gamma^i)$.
Note that the factors in the bilinears
of 4-component Dirac spinors
are `anti-symmetrized'.
The charge conjugation
parity of the bilinears
is easily determined
by the linearity
of the charge conjugation
transformation
of the spinor ${\mathsf \Psi}$
to be, respectively,
+1 (S), +1 (P), $-1$ (V),
+1 (A) and $-1$ (T).
From the expressions
for the bilinear covariants in terms of the spinor ${\mathsf \Psi}$, 
it is easy to see that
only 
(A) and (T) are invariant under the Pauli-G\"ursey transformations
${\mathsf \Psi}\to{\mathsf \Psi}q$ for $q\in SL(1,{\mathbb H})$.
This can be shown explicitly using the 4-component
notation.
}
Time-reversal in the one-particle
theory is defined by
${\mathsf \Psi}(x)\to{\mathsf \Psi}'(x_T)
=-\sigma_3{\mathsf \Psi}(x)e_2$,
which implies that
$\rho\big({\mathsf \Psi}'(x_T)\big)
=(\psi'(x_T), \Omega\psi{'}^*(x_T))
=\rho\big(-\sigma_3{\mathsf \Psi}(x)e_2\big)
=(-i\Omega\gamma^0\psi^*(x),
i\gamma^0\psi(x))$.
\\[2mm]
\ind
We are now in a position to
show that the transformation
${\mathsf \Psi}\to{\mathsf \Psi}q$ for $q\in SL(1,{\mathbb H})$
corresponds to the Pauli-G\"ursey transformation.
First, we define $\gamma_5=-i\gamma^0\gamma^1\gamma^2\gamma^3
=\left(
\barr{cc}
0_2&i1_2\\
-i1_2&0_2\\
\earr
\right)$\\
$=-i\rho(\omega)$,
so that
we have $\Omega \psi^*=-\gamma_5C{\tilde\psi}^T$,
leading to
\beq
\rho({\mathsf \Psi})
=(\psi, -\gamma_5C{\tilde\psi}^T).
\label{eqn:5-35}
\eeq
Similarly, we find
\beqa
\rho({\mathsf \Psi}q)
&=&(\psi, -\gamma_5C{\tilde\psi}^T)
\left(
\barr{cc}
q_0-iq_3&-iq_1-q_2\\
-iq_1+q_2&q_0+iq_3\\
\earr
\right)\nn\\[2mm]
&=&
(a\psi-b\gamma_5C{\tilde\psi}^T,
-b^*\psi-a^*\gamma_5C{\tilde\psi}^T),
\label{eqn:5-36}
\eeqa
where $a=q_0-iq_3$ and $b=-iq_1+q_2$, with
$|a|^2+|b|^2=1$ from the condition $|q|=1$.
Hence, we obtain
the result
that the transformation
${\mathsf \Psi}\to {\mathsf \Psi}'={\mathsf \Psi}q$
implies
\beqa
\psi'&=&a\psi-b\gamma_5C{\tilde\psi}^T,\nn\\[2mm]
{\tilde\psi}{'}^T&=&a^*{\tilde\psi}^T+b^*\gamma_5C\psi,
\label{eqn:5-37}
\eeqa
which is identical to the (canonical) Pauli-G\"ursey 
transformation. \cite{8)}
It is known that
the Pauli-G\"ursey group $SU(2)$
is the symmetry group of the
massless Dirac Lagrangian.
This fact is more transparent in our
quaternionic formalism.
\\[2mm]
\ind
To find the associated conserved current (in the massless case),
we consider the local infinitesimal Pauli-G\"ursey transformation,
$\delta{\mathsf \Psi}(x)={\mathsf \Psi}(x)e_a\delta q_a(x)$.
The variation of the kinetic energy part of the Dirac
Lagrangian (\ref{eqn:5-23}) is given by
\beq
\delta\cL_{m=0}=
i\mbox{Re}\,
[\delta{\mathsf \Psi}^\dag{\tilde\Gamma}^i
\partial_i{\mathsf \Psi}
+{\mathsf \Psi}^\dag{\tilde\Gamma}^i
\partial_i\delta{\mathsf \Psi}]
=
iJ_a^i\partial_i\delta q_a,
\label{eqn:5-38}
\eeq
where
\beq
J_a^i\equiv
\mbox{Re}\,[{\mathsf \Psi}^\dag
{\tilde\Gamma}^i
{\mathsf \Psi}e_a].
\label{eqn:5-39}
\eeq
Using (\ref{eqn:2-17})
we see
that
\beqa
J_1^i&=&
\frac 12[{\tilde\psi}\gamma^i\gamma_5\psi^c
+{\tilde\psi}^c\gamma^i\gamma_5\psi],\nn\\[2mm]
J_2^i&=&
\frac i2[{\tilde\psi}\gamma^i\gamma_5\psi^c
-{\tilde\psi}^c\gamma^i\gamma_5\psi],\nn\\[2mm]
J_3^i&=&
-\frac 12[{\tilde\psi}\gamma^i\psi
-{\tilde\psi}^c\gamma^i\psi^c].
\label{eqn:5-40}
\eeqa
The charges $Q_a=\int\,d^3\bx J_a^0(x)$
together form a closed set of commutation relations:
\beq
\big[\frac {Q_a}2, \frac {Q_b}2\big]
=-i\epsilon_{abc}\frac {Q_c}2.
\label{eqn:5-41}
\eeq
The minus sign
on the right-hand side here
stems from the fact that
the Pauli-G\"ursey transformation
${\mathsf \Psi}\to {\mathsf \Psi}q,$ with
$q\in SL(1,{\mathbb H})$,
is the right translation
rather than the left multiplication
as in the following example.
\\[2mm]
\ind
In terms of the decomposition (\ref{eqn:5-22}),
the Dirac Lagrangian (\ref{eqn:5-23})
becomes
\beq
\cL=i\mbox{Re}\,[{\mathsf \Psi}_+^\dag
{\tilde\Gamma}^i\partial_i
{\mathsf \Psi}_+-m{\mathsf \Psi}_+^\dag\sigma_3{\mathsf \Psi}_+e_3
+
{\mathsf \Psi}_-^\dag
{\tilde\Gamma}^i\partial_i
{\mathsf \Psi}_--m{\mathsf \Psi}_-^\dag\sigma_3{\mathsf \Psi}_-e_3].
\label{eqn:5-42}
\eeq
For the multi-component field
${\mathsf \Psi}_\pm$,
(\ref{eqn:5-42})
is invariant
under the transformation
\beq
{\mathsf \Psi}_+(x)
=g^{-1}{\mathsf \Psi}_+(x),\;\;\;
{\mathsf \Psi}_-(x)
=g^{*\,-1}{\mathsf \Psi}_-(x),\;\;\;
g\in SU(n),
\label{eqn:5-43}
\eeq
where we assume in this paragraph only that
${\mathsf \Psi}_+$ belongs to the
fundamental representation
of $SU(n)$,
so that ${\mathsf \Psi}_-$
transforms like its complex conjugate.
The associated conserved current
is obtained by considering the local
infinitesimasl transformations,
$\delta{\mathsf \Psi}_+(x)
=-iT_A\epsilon^A(x){\mathsf \Psi}_+(x)$
and $\delta{\mathsf \Psi}_-(x)
=iT_A^*\epsilon^A(x){\mathsf \Psi}_-(x)$,
with
$T_A$ ($A=1,2,\cdots, n^2-1$)
being the Hermitian generators
of $SU(n)$,
\beq
\delta\cL
=
(\partial_i\epsilon^A)\mbox{Re}\,
[{\mathsf \Psi}_+^\dag
{\tilde\Gamma}^i
T_A{\mathsf \Psi}_+
-{\mathsf \Psi}_-^\dag
{\tilde\Gamma}^i
T_A^*{\mathsf \Psi}_-]
=
iJ_A^i(\partial_i\epsilon^A),
\label{eqn:5-44}
\eeq
where
\beq
J_A^i
=-i\mbox{Re}\,
[{\mathsf \Psi}_+^\dag
{\tilde\Gamma}^i
T_A{\mathsf \Psi}_+
-{\mathsf \Psi}_-^\dag
{\tilde\Gamma}^i
T_A^*{\mathsf \Psi}_-]
=-
\frac 12[{\tilde\psi},
\gamma^iT_A\psi],
\label{eqn:5-45}
\eeq
again through application of (\ref{eqn:2-17}).
Their charges, $Q_A=\int\,d^3\bx J_A^0(x)$,
obey the commutation relations
\beq
[Q_A, Q_B]=if_{ABC}Q_C,
\label{eqn:5-46}
\eeq
where $f_{ABC}$ is the structure constant of $SU(n)$
defined by $[T_A, T_B]=if_{ABC}T_C$.
\\[2mm]
\ind
In passing, we remark that
the time-reversal in the second-quantized
theory is defined by
\beq
U_T{\mathsf \Psi}(x)U_T^{-1}=-\sigma_3{\bar
{\mathsf \Psi}(x_T)}e_2,
\label{eqn:5-47}
\eeq
only in the sense that
$U_T\rho\big({\mathsf \Psi}(x)\big)U_T^{-1}
=U_T\rho\big(-\omega{\bar{\mathsf \Psi}(x_T)}e_2\big)U_T^{-1}$
reproduces the usual time-reversal transformation,
$U_T\psi(x)U_T^{-1}
=R\psi(x_T)$ with $R^{-1}\gamma^{i\,*}R=
\gamma_i$.
\section{Complex quaternions and Dirac theory}                  %
As we have seen in the previous section,
the introduction of the ordinary imaginary unit $i$,
which commutes with Hamilton's units $\bi, \bj, \bk$,
into the theory
is not merely a notational technicality.
Rather, it is a necessary part of the theory
if we require the persistent
presence of the Pauli-G\"ursey
$SU(2)$ symmetry in the massless Dirac Lagrangian
based on the spinor group $Spin(2,{\mathbb H})$.
This allows us to introduce complex quaternions
${\mathsf \Psi}_\pm$ to quantize the theory.
We then assert that
no conceptual hurdle remains
to make full use of complex quaternions
to formulate the
spinor representation of the Lorentz group and the Dirac
theory. 
Of course, no novelty
is contained in this kind of reformulation.
Instead, it is intended to
show that complex quaternions
are useful mathematical tools
to formulate
the Dirac theory also 
in general relativity,
as shown in Ref. 14).
\\[2mm]
\ind
To determine the most natural
notational usefulness of complex quaternions 
for the description
of the
spinor representation of the restricted Lorentz group and the Dirac theory,
we recall that
the Dirac representation
is a direct sum of two irreducible
spinor representations, $D^{(1/2,0)}\oplus
D^{(0,1/2)}$,
which is reflected in (\ref{eqn:5-4}).
What does this mean in our quaternionic
formulation using $Spin(2,{\mathbb H})$?
The answer is simple:
There must be a unitary transformation
which {\it diagonalizes} the Lorentz matrix
(\ref{eqn:4-5}) and the $\Gamma^i$ matrix. 
It is important to recognize
that, for our purpose, there is nothing  preventing
use
of the ordinary 
imaginary unit $i$
commuting with Hamilton's units.
Hence, we have the freedom to
define the unitary matrix
over the complex field ${\mathbb C}$.
In fact, the unitary matrix
\beq
S=\frac 1{\sqrt{2}}
  \left(
  \barr{cc}
  1&-i\\
  1&i\\
  \earr
  \right)
\label{eqn:6-1}
\eeq
can be used for this purpose, as we have
\beq
S{\mathsf A}S^{-1}=
  \left(
  \barr{cc}
  U&0\\
  0&U^*\\
  \earr
  \right),
S\Gamma^i S^{-1}=
  \left(
  \barr{cc}
  -b^i&0\\
  0&-{\bar b}^i\\
  \earr
  \right),
S{\tilde\Gamma}^i S^{-1}=
  \left(
  \barr{cc}
  -{\bar b}^i&0\\
  0&-b^i\\
  \earr
  \right),
\label{eqn:6-2}
\eeq
where we have defined
\beqa
U&\equiv& Q-iP,\;\;\;  U{\bar U}=1,\\[2mm]
b_0&\equiv& e_0,\;\;\;
b_1\equiv ie_1,\;\;\;
b_2\equiv ie_2,\;\;\;
b_3\equiv ie_3,\;\;\;
b^i=\eta^{ij}b_j.
\label{eqn:6-3}
\eeqa
The set $\{b_0, b_1, b_2, b_3\}$
forms a basis of complex quaternions, ${\mathbb H}^c$,
with
\beq
b_i{\bar b}_j
+b_j{\bar b}_i=-2\eta_{ij},\;\;\;
{\bar b}_ib_j
+{\bar b}_jb_i=-2\eta_{ij}.
\label{eqn:6-5}
\eeq
We express a complex quaternion $q\in{\mathbb H}^c$ in
a covariant way
as
\beq
q=b_0q^0+b_1q^1+b_2q^2+b_3q^3
\equiv b_iq^i,\;\;\;
q^i\in{\mathbb C}.
\label{eqn:6-6}
\eeq
There are three operations,
quaternionic conjugation, $^-$,
complex conjugation, $^*$, and Hermitian
conjugation, $^\dag$, for a complex quaternion $q\in{\mathbb H}^c$:
\beqa
{\bar q}&=&b_0q^0-b_1q^1-b_2q^2-b_3q^3,\nn\\[2mm]
q^*&=&b_0(q^0)^*-b_1(q^1)^*
-b_2(q^2)^*-b_3(q^3)^*,\nn\\[2mm]
q^\dag&=&b_0(q^0)^\dag+b_1(q^1)^\dag
+b_2(q^2)^\dag+b_3(q^3)^\dag.
\label{eqn:6-7}
\eeqa
Here, $(q^i)^\dag$ represents the usual
Hermitian conjugation.
Nonethelesss,
we write $q^*={\bar q}^\dag$
for both complex numbers and operators $q^i$.
For $p, q\in{\mathbb H}^c,$ we have
\beqa
\overline{pq}&=&{\bar q}{\bar p},\\[2mm]
(pq)^\dag&=&q^\dag p^\dag.
\label{eqn:6-8}
\eeqa
The real and imaginary parts of $q\in{\mathbb H}^c$ are given,
respectively, by
\begin{eqnarray}
\mbox{Re}\,q\,=
\frac 12(q+{\bar q})
=\mbox{Re}\;\,{\bar q}\,,\hspace{1cm}
\mbox{Im}\,q\,=\frac 12(q-{\bar q})
=-\mbox{Im}\,{\bar q}\,.
\label{eqn:6-10}
\end{eqnarray}
Since this is the same as the definition
for the algebra ${\mathbb H}$,
$\mbox{Re}\,q\,$ for $q\in{\mathbb H}^c$
is, in general, complex.
The real part of the product of two factors
is independent of the order of the factors,
as in (\ref{eqn:2-5}):
\beq
\mbox{Re}\,[qp]
=\mbox{Re}\,[pq].
\label{eqn:6-11}
\eeq
This leads to the
cyclic property of
the operation Re$\,[\cdots]$
of an arbitrary number of factors.
The norm of a complex quaternion $q\in{\mathbb H}^c$
is no longer positive definite:  
\beq
N(\,q\,)=q{\bar q}={\bar q}q\equiv
|q|^2=(q^0)^2-(q^1)^2-(q^2)^2-(q^3)^2
\equiv -\eta_{ij}q^iq^j.
\label{eqn:6-12}
\eeq
Thus,
although $N(qp)=N(q)N(p)$ for $q, p\in{\mathbb H}^c$
still holds, the algebra ${\mathbb H}^c$ is not a division algebra.
Unit complex quaternions form a group,
which we write
\beq
SL(1,{\mathbb H}^c)=\{q\in{\mathbb H}^c;\;|q|=1\}.
\label{eqn:6-13}
\eeq
By definition,
$U$ of (6$\cdot$3)
belongs to this group, $U\in SL(1,{\mathbb H}^c)$.
The Lorentz transformation
(\ref{eqn:4-9})
becomes
\beq
{\mathsf x}\to {\mathsf x}'=U{\mathsf x}U^\dag,
\label{eqn:6-14}
\eeq
where ${\mathsf x}=b_ix^i$ [see also
(\ref{eqn:1-6}) and (\ref{eqn:1-7})],
with
$|{\mathsf x}|^2=-\eta_{ij}x^ix^j$. Writing
$x{'}^i=\Lambda^i_{\;\;j}x^j$,
we obtain
\beq
\Lambda^i_{\;\;j}b_i=Ub_jU^\dag,
\label{eqn:6-15}
\eeq
which determines a continuous function
$\Lambda^i_{\;\;j}=\Lambda^i_{\;\;j}(U)$.
Together with its quaternionic conjugation
and (6$\cdot$5),
this equation leads to the condition
$\Lambda^T\eta\Lambda=\eta$.
Since
Re$\,[b_i{\bar b}_k]=-\eta_{ik}$ from (6$\cdot$5),
multiplying (\ref{eqn:6-15})
by ${\bar b}^k$ from right
and taking the real part,
we obtain 
\beq
\Lambda^i_{\;\;j}=-\mbox{Re}\,[Ub_jU^\dag{\bar b}^i].
\label{eqn:6-16}
\eeq
This gives
$\Lambda^0_{\;\;0}=\mbox{Re}\,[UU^\dag]
=\mbox{Re}\,[(Q-iP)({\bar Q}+i{\bar P})]
=|Q|^2+|P|^2=|Q|^2-|P|^2+2|P|^2=
1+2|P|^2\ge 1$.
We also have
\beq
\Lambda^i_{\;\;m}b_i
\Lambda^j_{\;\;n}{\bar b}_j
\Lambda^k_{\;\;s}b_k
\Lambda^l_{\;\;t}{\bar b}_l
=(Ub_mU^\dag)({\bar U}^\dag{\bar b}_n{\bar U)
(Ub_sU^\dag)({\bar U}^\dag{\bar b}_t{\bar U})
=Ub_m{\bar b}_nb_s{\bar b}_t\bar U}.
\label{eqn:6-17}
\eeq
Taking the real part of this equation
and using the relation
\beq
\mbox{Re}\,[b_i{\bar b}_jb_k{\bar b}_l]
=\eta_{ij}\eta_{kl}-\eta_{ik}\eta_{jl}+\eta_{il}\eta_{jk}
-i\epsilon_{ijkl},
\label{eqn:6-18}
\eeq
where $\epsilon_{ijkl}$ is the 4-dimensional
Levi-Civita symbol with
$\epsilon_{0123}=-1$,
we obtain (since $\eta_{ij}\Lambda^i_{\;\;k}
\Lambda^j_{\;\;l}=\eta_{jl}$)
\beq
\Lambda^i_{\;\;m}
\Lambda^j_{\;\;n}
\Lambda^k_{\;\;s}
\Lambda^l_{\;\;t}\epsilon_{ijkl}
=\epsilon_{mnst}.
\label{eqn:6-19}
\eeq
Thus we find det$\,\Lambda=1$.
Moreover, if we write
$\Lambda^i_{\;\;j}(U_1)
=-\mbox{Re}\,[\alpha^ib_j]$
and
$\Lambda^j_{\;\;k}(U_2)
=-\mbox{Re}\,[{\bar b}^j\beta_k]$
with
$\alpha^i=U_1^\dag{\bar b}^iU_1$
and
$\beta_k=U_2b_kU_2^\dag$,
we can easily compute
\beq
\Lambda^i_{\;\;j}(U_1)\Lambda^j_{\;\;k}(U_2)
=\frac 14
[b_j(\alpha^i\beta_k){\bar b}^j
+
(b_j\alpha^ib^j){\bar\beta}_k
+
{\bar \alpha}^i({\bar b}_j\beta_k{\bar b}^j)
+
{\bar \alpha}^i({\bar b}_jb^j){\bar \beta}_k
].
\label{eqn:6-20}
\eeq
Then, recalling the relations
$b_j{\bar b}^j=-4,
b_jq{\bar b}^j=-2(q+{\bar q})$ and
$b_jqb^j=2{\bar q}$, we arrive at the homomorphism
\beq
\Lambda^i_{\;\;j}(U_1)\Lambda^j_{\;\;k}(U_2)
=-\mbox{Re}\,[\alpha^i\beta_k]
=
\Lambda^i_{\;\;k}(U_1U_2).
\label{eqn:6-21}
\eeq
Thus we have established the isomorphism\footnote{The
proof is a mere repetition of that in \S 4 
using the diagonalization
(\ref{eqn:6-2}), but it is presented above
in order to allow comparison with a similar proof in
spinor analysis.}
\beq
SL(1,{\mathbb H}^c)/Z_2{\cong}L_+^\uparrow.
\label{eqn:6-22}
\eeq
Noting that
the representation $\rho$
of the unit complex quaternion $U$
is given by
\beq
\rho(U)
=
\left(
\barr{cc}
U^0+U^3&U^1-iU^2\\
U^1+iU^2&U^0-U^3\\
\earr
\right)
=\sigma_iU^i\in SL(2,{\mathbb C}),
\label{eqn:6-23}
\eeq
we conclude that
\beq
SL(1,{\mathbb H}^c)\cong SL(2,{\mathbb C}).
\label{eqn:6-24}
\eeq
We have thus
proved the well-known isomorphism
(\ref{eqn:5-1}) within the terminology
of complex quaternions
inheritated from $Spin(2,{\mathbb H})/Z_2\cong L_+^\uparrow$.
In particular,
(\ref{eqn:6-14})
becomes
the well-known transformation
$X'\to
X=AXA^\dag$,
where $X$ is given by (\ref{eqn:1-1})
and $A=\rho(U)$.
\\[2mm]
\ind
To express the unit quaternion
$U$ in terms of the Lorentz matrix $\Lambda^i_{\;\;j}$,
we 
multiply (\ref{eqn:6-15}) by ${\bar b}^j$ from right
and sum over $j$ to obtain
\beq
U=
\pm \frac{\Lambda^i_{\;\;j}b_i{\bar b}^j}
{|\Lambda^i_{\;\;j}b_i{\bar b}^j|},
\label{eqn:6-25}
\eeq
where we have again used the relation
$b_jq{\bar b}^j=-2(q+{\bar q})$.
This formula is useful to
obtain the infinitesimal generators
in the spinor representation.
Thus, for infinitesimal transformations
$\Lambda^i_{\;\;j}
=\eta^i_{\;\;j}
+\omega^i_{\;\;j}$,
the corredsponding unit complex quaternion
is given by
$\pm U=1-
\frac 18\omega_{ij}(b^i{\bar b}^j-b^j{\bar b}^i)$.
This determines the Lorentz generators,
$i(b^i{\bar b}^j-b^j{\bar b}^i)/4$,
for the spinor-quaternion defined below.
\\[2mm]
\ind
For the $Spin(2,{\mathbb H})$ Dirac spinor,
we have
\beq
S{\mathsf \Psi}=
\left(
\barr{l}
\psi\\
\psi^*\\
\earr
\right),\;\;\;
\left\{
\barr{l}
\psi=\frac 1{\sqrt{2}}(\zeta^1-i\zeta^2)\equiv b_i\psi^i,\\[2mm]
\psi^*=\frac 1{\sqrt{2}}(\zeta^1+i\zeta^2)
\equiv (b_i)^*(\psi^i)^*={\bar \psi}^\dag.\\
\earr
\right.
\label{eqn:6-26}
\eeq
Here and hereafter,
we use the notation $\psi$
for the spinor-quaternion $\psi=b_i\psi^i$.\footnote{The Majorana
spinor-quaternion
satisfies the condition
$\psi b_1=\psi$,
because $\zeta^2=-\zeta^1e_1$
leads to
$\psi=(1/\sqrt{2})(\zeta^1+i\zeta^1e_1)
=(1/\sqrt{2})\zeta^1(1+b_1)$ and $b_1^2=1$.}
It is a complex Grassmannian quaternion.
Let $q$ and $p$ be any Grassmannian complex quaternions.
The formulae (\ref{eqn:6-11}),
(6$\cdot$8) and
(\ref{eqn:6-12}), valid for ordinary complex quaternions,
should be changed as
\beqa
\mbox{Re}\,[qp]
&=&-\mbox{Re}\,[pq],\nn\\[2mm]
\overline{qp}
&=&-{\bar p}{\bar q},\nn\\[2mm]
N(q)&=&q{\bar q}
=e_a(q_aq_0-q_0q_a)
-\epsilon_{abc}q^aq^be_c,
\label{eqn:6-27}
\eeqa
the first two of which
were already used in the previous section, in
(\ref{eqn:5-24}).
The third equation implies that
$\mbox{Re}[\,|q|^2\,]=0$ for a Grassmannian complex quaternion $q$.
The spinor-quaternion
is subject to the Lorentz transformation
\beq
\psi(x)\to
\psi'(x')=U\psi(x),\;\;\;
\psi^*(x)\to
\psi{'}^*(x')=U^*\psi^*(x).
\label{eqn:6-28}
\eeq
Componentwise, we have 
\beq
\psi{'}^j(x')
=\big(\omega(b_iU^i)\psi\big)^j(x)
=(\Omega_iU^i\psi)^j(x),\;\;\;
(i,j=0,1,2,3),
\label{eqn:6-29}
\eeq
where
the matrices
$\Omega_i$ are defined by
(\ref{eqn:2-21}):
\beq
\omega(b_i)=
\Omega_i.\;\;\;(i=0,1,2,3)
\label{eqn:6-30}
\eeq
\ind
The spinor representation
(\ref{eqn:6-29}) 
is 4-dimensional
and decomposable
into
two
2-dimensional equivalent spinor representations,
as noted in (\ref{eqn:2-22}).
This means that
we have the reduced transformation law
\beq
(V\psi){'}^i(x')
=
\left(
\barr{cc}
\rho(U)&0\\
0&{\bar \rho}(U)\\
\earr
\right)_{ij}
(V\psi)^j(x),\;\;\;
(i,j=0,1,2,3)
\label{eqn:6-31}
\eeq
where $V$ is given by (\ref{eqn:2-22}).
In other words,
if we set
\beqa
\xi^1&=&(V\psi)^0=\psi^0+\psi^3,\;\;\;
\xi^2=(V\psi)^1=\psi^1+i\psi^2,\nn\\[2mm]
\eta_1&=&(V\psi)^2=\psi^0-\psi^3,\;\;\;
\eta_2=(V\psi)^3=-\psi^1+i\psi^2,
\label{eqn:6-32}
\eeqa
we find that
\beqa
\xi{'}^\alpha(x')&=&A^\alpha_{\;\;\beta}\xi^\beta(x),\;\;\;
\eta{'}_\alpha(x')=A_\alpha^{\;\;\beta}\eta_\beta(x),\;\;\;
(\alpha, \beta=1,2)
\label{eqn:6-33}
\eeqa
where
$\big(\rho(U)\big)_{\alpha\beta}
=(\sigma_iU^i)_{\alpha\beta}
\equiv A^\alpha_{\;\;\beta}$
and
$\big({\bar \rho}(U)\big)_{\alpha\beta}
=({\tilde\sigma}_i^*U^i)_{\alpha\beta}
\equiv A_\alpha^{\;\;\beta}$.\footnote{Inspection shows that
$\big({\bar\rho}(U)\big)^{T\,-1}
=\rho(U)$, since det$\,\rho(U)=1$ from $|U|=1$.
The relation (\ref{eqn:2-18}) is
still valid for a complex quaternion,
$\omega{\bar\rho}(U)\omega^{-1}
=\rho(U)$, and thus
the
representation ${\bar\rho}(U)$ is equivalent to
$\rho(U)$, because det$\,\omega=1$.
Note, however, that
${\bar\rho}(U)$ is not equal to
$\rho^*(U)$.
Therefore, the spinor-quaternion
contains only Weyl spinors
with undotted indices,
while its complex conjugate
contains only those with dotted indices.}\
To go further, we set
\beqa
v_1&=&(bV^{-1})_0=\frac 12(b_0+b_3),\;\;\;
v_2=(bV^{-1})_1=\frac 12(b_1-ib_2),\nn\\[2mm]
v_1^*&=&(bV^{-1})_2=\frac 12(b_0-b_3),\;\;\;
v_2^*=(bV^{-1})_3=-\frac 12(b_1+ib_2),
\label{eqn:6-34}
\eeqa
and therefore
we obtain
\beq
\psi=b_i\psi^i
=(bV^{-1})_i(V\psi)^i
=v_\alpha\xi^\alpha
+v_\alpha^*\eta_\alpha.
\label{eqn:6-35}
\eeq
This is equivalent to
the decomposition
proposed by Rastall \cite{2)},
using the projection operators
$p_\pm=(b_0\pm b_3)/2$,
\beqa
\psi&=&\psi(p_++p_-)
\equiv\psi_L+\psi_R,\nn\\[2mm]
\psi_L&\equiv&\psi p_+=v_\alpha\xi^\alpha,\;\;\;
\psi_R\equiv\psi p_-=v_\alpha^*\eta_\alpha,
\label{eqn:6-36}
\eeqa
where
$v_1=b_0p_+=b_3p_+,
v_2=b_1p_+=-ib_2p_+,
v_1^*=b_0p_-=-b_3p_-,
v_2^*=-b_1p_-=-ib_2p_-$.
The basis $\{v_1, v_2, v^*_1, v^*_2\}$
is called the split basis,
and the decomposition (\ref{eqn:6-36})
is called the chiral decomposition in Ref. 14).
Note that
the spinor-quaternion
$\psi$  
has two 2-component spinors with
undotted indices only,
while its complex conjugate $\psi^*$
has two 2-component spinors with
dotted indices only.\footnote{The Majorana condition
$\psi b_1=\psi$
implies $\xi^\alpha=\eta^\alpha$,
where we define $\eta^\alpha=(\omega)_{\alpha\beta}\eta_\beta$.
The Majorana spinor is given by
$\psi_L$ only as
$\psi^M=\psi_L+i\psi_Lb_2$.
(There is an error in the sign
of this equation in Ref. 14).)}
\\[2mm]
\ind
The Dirac Lagrangian
(\ref{eqn:5-23})
becomes
\beq
\cL=i\mbox{Re}\,\Biggl[(\psi^\dag, {\bar\psi})
\left(
\barr{cc}
{\bar \partial}&0\\
0&\partial\\
\earr
\right)
\left(
\barr{l}
\psi\\
{\bar\psi}^\dag\\
\earr
\right)
+im
(\psi^\dag, {\bar\psi})
\left(
\barr{l}
{\bar\psi}^\dag\\
\psi\\
\earr
\right) b_3\Biggr],
\label{eqn:6-37}
\eeq
where $\partial=-b^i\partial_i$.
The first term of this Lagrangian
has the symmetry
$\psi\to\psi q$ for $q\in SL(1,{\mathbb H})$,
which induces
${\bar\psi}^\dag\to
({\overline{\psi q}})^\dag=({\bar q}{\bar\psi})^\dag
={\bar\psi}^\dag q$. 
This is merely the Pauli-G\"ursey symmetry.
Although this symmetry is hidden,
we are led to redefine
\beq
\left(
\barr{l}
\psi\\
{\bar\psi}^\dag\\
\earr
\right)
={\mathsf \Psi}_1+
{\mathsf \Psi}_2,\;\;\;
{\mathsf \Psi}_1\equiv
\left(
\barr{l}
\psi_L\\
{\bar\psi}_R^\dag\\
\earr
\right),\;\;\;
{\mathsf \Psi}_2\equiv
\left(
\barr{l}
\psi_R\\
{\bar\psi}_L^\dag\\
\earr
\right),
\label{eqn:6-38}
\eeq
to cast (\ref{eqn:6-37})
in the form
\beq
\cL=
i\mbox{Re}\,\big[
{\mathsf \Psi}_1^\dag
\left(
\barr{cc}
{\bar \partial}&im\\
im&\partial\\
\earr
\right){\mathsf \Psi}_1
+{\mathsf \Psi}_2^\dag
\left(
\barr{cc}
{\bar \partial}&-im\\
-im&\partial\\
\earr
\right){\mathsf \Psi}_2\big].
\label{eqn:6-39}
\eeq
In this form, the transformation
$\psi_L\to\psi_Lq$
never implies ${\bar\psi}_R^\dag q$
for any $q\in SL(1,{\mathbb H})$.
Only in the massless case does
the spinor ${\mathsf \Psi}_1+{\mathsf \Psi}_2$
determine the Lagrangian, so that
the transformation $({\mathsf \Psi}_1+{\mathsf \Psi}_2)\to
({\mathsf \Psi}_1+{\mathsf \Psi}_2)q$ is allowed for
$q\in SL(1,{\mathbb H})$.
\\[2mm]
\ind
The Lagrangian (\ref{eqn:6-39})
was utilized to incorporate the Higgs mechanism
into the theory in Ref.~5).
Here we instead make explicit
the spinor (undotted and dotted)
indices of the spinor-quaternion.
Using the relations
\beq
b^iv_\alpha=v_\beta(\sigma^i)_{\beta\alpha},\;\;\;
b^iv_\alpha^*=v_\beta^*({\tilde\sigma}^{i\,*})
_{\beta\alpha},\;\;\;
{\bar b}^iv_\alpha=v_\beta({\tilde\sigma}^i)_{\beta\alpha},
\;\;\;
{\bar b}^iv_\alpha^*=v_\beta^*(\sigma^{i\,*})
_{\beta\alpha},
\label{eqn:6-40}
\eeq
we directly verify the transformation properties
(omitting the arguments $x$ and $x'$)
\beqa
U\psi_L&=&v_\alpha A^\alpha_{\;\;\beta}\xi^\beta=\psi'_L
=v_\alpha \xi{'}^\alpha,\nn\\[2mm]
U\psi_R&=&v_\alpha^*A_\alpha^{\;\;\beta}\eta_\beta=\psi'_R
=v_\alpha^*\eta{'}_\alpha,\nn\\[2mm]
U^*{\bar\psi}_L^\dag&=&
v_\alpha^* A^{\dot\alpha}_{\;\;{\dot\beta}}{\bar\xi}^{\dot\beta}
={\bar\psi}{'}_L^\dag
=v_\alpha^* {\bar\xi}{'}^{\dot\alpha},\nn\\[2mm]
U^*{\bar\psi}_R^\dag&=&v_\alpha A_{\dot\alpha}^{\;\;{\dot\beta}}
{\bar\eta}_{\dot\beta}=
{\bar\psi}{'}_R^\dag
=v_\alpha {\bar\eta}{'}_{\dot\alpha}.
\label{eqn:6-41}
\eeqa
The first two equations
here
are identical to
(\ref{eqn:6-33}),
and the last two
contain the matrix elements
$A^{\dot\alpha}_{\;\;{\dot\beta}}
=(A^\alpha_{\;\;\beta})^*$
and
$A_{\dot\alpha}^{\;\;{\dot\beta}}
=(A_\alpha^{\;\;\beta})^*$.
We thus recover the well-known
Lorentz transformation property
of the Weyl spinors,
as indicated in (\ref{eqn:5-3}) and (\ref{eqn:5-4}). 
This result is reflected by a left-translation
$\psi\to U\psi$, and the chiral decomposition (\ref{eqn:6-36}),
which, incidentally, carry no spinor indices.
If we regard the spinor-quaternion
as a $2\times 2$ matrix field via the Pauli representation,
it should not be forgotten
that
the matrix field
represents 2 (independent) Wely spinors with undotted indices,
while
the complex conjugate
of the matrix field involves 2 (independent)
Wely spinors with dotted indices.
\section{Conclusions}                                          %
As a continuation
of the previous works,
we have shown the existence of the
spinor group $Spin(2,{\mathbb H})$
over the field of real quaternions,
which is locally isomorphic to
the restricted Lorentz group.
This establishes a close connection
between real quaternions and our space-time,
and it
helps reformulate the Dirac theory
such that
the factors
in the Lagrangian
and the bilinear covariants
are automatically `anti-symmetrized'
upon using the Pauli representation
of quaternions,
which involves $i$.
This is in conformity with 
our refomulation,
which requires an additional
imaginary unit
commuting with Hamilton's units,
leading to
the scheme
of complex quaternions.
In contrast to
the usual wisdom
that
complex quaternions
are formally connected
to relativity
through
the imaginary time coordinate,
the present way
of introducing
$i$, which commutes with Hamilton's units,
depends on a consistent
Lagrangian
formulation
of the Dirac theory using $Spin(2,{\mathbb H})$.
It is not connected with
the imaginary time coordinate,
because the quaternionic Dirac equation
(\ref{eqn:5-10}),
though relativistic,
does not allow the presence of $i$
by construction,
and the Pauli-G\"ursey symmetry
for the massless case
never requires $i$ at the level of
the equation.
\\[2mm]
\ind
To close this paper,
we again emphasize 
the fact that
quaternions and the space-time have the
same dimensionality,
in relation
with general relativity.
Let us rewrite the Dirac Lagrangian
(\ref{eqn:6-37})
in the form
\bea*
\cL=
i\mbox{Re}\,[\psi^\dag{\bar \partial}\psi
+{\bar\psi}\partial{\bar\psi}^\dag]
+\mbox{mass term}.
\eea*
This is Lorentz invariant by construction.
To make it invariant
under the general
coordinate transformations
$x^\mu\to
x{'}^\mu=x{'}^\mu(x)\;(\mu=0,1,2,3)$,
we have to assume, as usual, that
the spinor-quaternion is a scalar,
\bea*
\psi'(x')=\psi(x),
\eea*
because it
has a single component,
and replace
the gradient
$\partial=-b^i\partial_i$
with
\bea*
D=-b^\mu\partial_\mu,
\eea*
where the vector-quaternion 
$b^\mu=b^\mu(x)$
transforms as
\bea*
b{'}^\mu(x')
=\frac{\partial x{'}^\mu}
{\partial x^\nu}b^\nu(x).
\eea*
Since $b^\mu(x)$ is a Hermitian quaternion,
we can write it
as
\bea*
b^\mu(x)=b^ib_i^{\;\;\mu}(x),
\eea*
where the vierbein $b_i^{\;\;\mu}$
is real field.
It then follows from
(\ref{eqn:6-5})
that
\bea*
b^\mu(x){\bar b}^\nu(x)
+b^\nu(x){\bar b}^\mu
=-2g^{\mu\nu}(x),
\eea*
where
\bea*
g^{\mu\nu}(x)=\eta^{ij}b_i^{\;\;\mu}(x)b_j^{\;\;\nu}(x).
\eea*
This means that we can regard
$g_{\mu\nu}$ (the inverse of $g^{\mu\nu}$)
as a metric tensor in general relativity.
If we also require the local Lorentz invariance
of the Dirac Lagrangian,
the quaternionic nature
of the
spinor quaternion
introduces a coupling of it to a
Yang-Mills type
gauge field 
$A_\mu=(1/8)A_{\mu ij}(b^i{\bar b}^j-b^j{\bar b}^i)$,
which has 6 components for each $\mu$,
since the Lorentz group is a 6-parameter group.
The final result is the Poincar\'e gauge theory
with the metric condition
yielding the 
Riemann-Cartan space-time.
This ends our brief comments
on the application
of the formalism
in \S 6, which is thoroughly discussed in Ref. 14). 
\appendix
\section{Some Propertis of $Spin(2,{\mathbb H})$ Matrices}                %
The matrix (\ref{eqn:4-5}) of $Spin(2,{\mathbb H})$
has several interesting properties,
which are given in Ref. 4) without proofs.
Although they are not all needed in the analysis given in the main text,
here we list them with proofs
included for completeness.
\beqa
&&\mbox{(I)}\;\;\;
{\mathsf A}=\left(
\barr{cc}
Q&-P\\
P&Q\\
\earr
\right)
=e^{B(q,p)},\;\;\;
B(q,p)=
\left(
\barr{cc}
q&-p\\
p&q\\
\earr
\right),\;\;\;q,p\in{\mathbb H}_0.\\[2mm]
&&\mbox{(II)}\;\;\;
{\mathsf A}=\left(
\barr{cc}
Q&-P\\
P&Q\\
\earr
\right)
=R(a)B(\mu)=B(\mu')R(a),\\[2mm]
&&\hspace{1cm}
R(a)=\left(
\barr{cc}
a&0\\
0&a\\
\earr
\right),\;\;\;a=\frac Q{|Q|},\\[2mm]
&&\hspace{1cm}
B(\mu)=\frac 1{\sqrt{1+\mu^2}}
\left(
\barr{cc}
1&\mu\\
-\mu&1\\
\earr
\right),\hspace{0.32cm}
\mu=-Q^{-1}P\in{\mathbb H}_0,\;\;\;
\mu'=-PQ^{-1}\in{\mathbb H}_0.\\[2mm]
&&\mbox{(III)}\;\;\;
R(a)=e^{B(q, 0)},\;\;\;a=e^q,\\[2mm]
&&\hspace{1cm}
B(\mu)=e^{B(0,p)},\;\;\;\mu=-\frac p{|p|}\tanh{|p|}.\\[2mm]
&&\mbox{(IV)}\;\;\;
R(a)\;\mbox{describes spatial rotation},\;
\mbox{while}\;B(\mu)\;\mbox{represents boost}.\nn
\label{eqn:A-1}
\eeqa
Here, ${\mathbb H}_0$ is the set of pure quaternions,
${\mathbb H}_0=\{q\in{\mathbb H};\;{\bar q}=-q\}$.
\\[2mm]
\ind
The proof of (A$\cdot$1) is by induction.
Assume that the 
$n$-th term in the sum
\bea*
e^{B(q,p)}
=\sum_{n=0}^\infty
\frac 1{n!}B^n(q,p)
\eea*
is of the form ${\mathsf A}$:
\bea*
n\mbox{-th term in}\;
e^{B(q,p)}
\equiv\frac 1{n!}B^n(q,p)
=
\left(
\barr{cc}
Q'&-P'\\
P'&Q'\\
\earr
\right).
\eea*
Then the $(n+1)$-st term
is
\bea*
\left(
\barr{cc}
Q'&-P'\\
P'&Q'\\
\earr
\right)
\left(
\barr{cc}
q&-p\\
p&q\\
\earr
\right)
=
\left(
\barr{cc}
Q'q-P'p&-Q'p-P'q\\
P'q+Q'p&-P'p+Q'q\\
\earr
\right),
\eea*
which is also of the form
${\mathsf A}$.
Since the first term $B(q,p)$
is already of the
form ${\mathsf A}$,
the sum $\sum_{n=0}^\infty
\frac 1{n!}B^n(q,p)$
is also of the form ${\mathsf A}$.
Next, we have ${\bar {\mathsf A}}{\mathsf A}
=e^{{\bar B}}e^{B}
=e^{-B}e^{B}=1_2$, because $B$ has only pure quaternions.
\\[2mm]
\ind
The second equality in (A$\cdot$2)
is obtained by writing
$P=-Q\mu$, and 
the third by putting
$P=-\mu'Q$.
Thus we have
$1+\mu^2=1-|\mu|^2
=1-|Q^{-1}P|^2=|Q|^{-2}(|Q|^2-|P|^2)
=|Q|^{-2}$
and simlarly
$1+\mu{'}^2=|Q|^{-2}$.
\\[2mm]
\ind
To prove (A$\cdot$5),
we note that
\bea*
e^{B(q,0)}
&=&
\left(
\barr{cc}
1&0\\
0&1\\
\earr
\right)
+\left(
\barr{cc}
q&0\\
0&q\\
\earr
\right)
+
\frac 1{2!}
\left(
\barr{cc}
q^2&0\\
0&q^2\\
\earr
\right)
+\cdots
=
\left(
\barr{cc}
e^q&0\\
0&e^q\\
\earr
\right)
=
\left(
\barr{cc}
a&0\\
0&a\\
\earr
\right),
\eea*
with $a=e^q$ being a unit quaternion,
because $q\in{\mathbb H}_0$.
To prove (A$\cdot$6)
we also use the expansion
\bea*
e^{B(0,p)}
&=&
\left(
\barr{cc}
1&0\\
0&1\\
\earr
\right)
+\left(
\barr{cc}
0&-p\\
p&0\\
\earr
\right)
+
\frac 1{2!}
\left(
\barr{cc}
-p^2&0\\
0&-p^2\\
\earr
\right)
+\cdots\\[2mm]
&=&
\left(
\barr{cc}
\cosh{|p|}&-(p/|p|)\sinh{|p|}\\
(p/|p|)\sinh{|p|}&\cosh{|p|}\\
\earr
\right)
\equiv
\frac 1{\sqrt{1+\mu^2}}
\left(
\barr{cc}
1&\mu\\
-\mu&1\\
\earr
\right),
\eea*
with
$\mu=-(p/|p|)\tanh{|p|}$
and
$1/\sqrt{1+\mu^2}=\cosh{|p|}$.
\\[2mm]
\ind
The property (IV)
is checked as follows.
For $R(a)=\left(
\barr{cc}
a&0\\
0&a\\
\earr
\right)$, we have
\bea*
{\mathsf X}'
=R(a){\mathsf X}R^\dag(a)
=x^0\Gamma_0+
\left(
\barr{cc}
0&-a(\bx\cdot\be)a^{-1}\\
a(\bx\cdot\be)a^{-1}&0\\
\earr
\right).
\eea*
This implies
that
$x{'}^0=x^0,
x{'}^a=
R^a_{\;\;b}x^b\;(a,b=1,2,3)$
with
$R^a_{\;\;b}
=n_an_b+(\delta_{ab}-n_an_b)\cos{\theta}
-\epsilon_{abc}n_c\sin{\theta}$
if
$a
=\cos{(\theta/2)}+e_an_a\sin{(\theta/2)}$,
which follows 
from $q=e_an_a(\theta/2)$.\footnote{The suffix $a$ 
should not be confused with the pure quaternion $a$.}
We thus find that
$R(a)$ describes spatial rotations
about the direction specified by the unit
vetor $\bn=(n_1,n_2,n_3)$.
For
$B(\mu)=
(1/\sqrt{1+\mu^2})
\left(
\barr{cc}
1&\mu\\
-\mu&1\\
\earr
\right)=
\exp{\left(
\barr{cc}
0&-p\\
p&0\\
\earr
\right)}$
with $\mu=e_an_a\tanh{(\zeta/2)}$
for $p=-e_an_a(\zeta/2)$,
we have, from ${\mathsf X}'
=B(\mu){\mathsf X}B^\dag(\mu)$,
\bea*
x{'}^0&=&
\frac {1-\mu^2}{1+\mu^2}x^0
-\frac 1{1+\mu^2}(\mu(\bx\cdot\be)
+(\bx\cdot\be)\mu)
=(\cosh{\zeta})\,x^0+(\sinh{\zeta})\,(\bn\cdot\bx),\\[2mm]
\bx'\cdot\be&=&
\frac 1{1+\mu^2}(\bx\cdot\be-
\mu(\bx\cdot\be)\mu+2\mu\,x^0)
=
\Big[
\bx
+(\cosh{\zeta}-1)\bn(\bn\cdot\bx)+(\sinh{\zeta})x^0\bn\Big]\cdot\be.
\eea*
Hence, we conclude
that
$B(\mu)$
describes a boost along the direction $\bn$.

\end{document}